\documentclass[aps,reprint,prl]{revtex4-1}
\usepackage{amsfonts, amsmath, amssymb, wasysym}
\usepackage{changes}
\usepackage[pdftex]{graphicx}

\newcommand{\vsigma}{{\vec\sigma}}
\newcommand{\XX}{{\mathcal X}}
\newcommand{\CC}{{\mathcal C}}

\newcommand{\const}{{\mathrm{const}}}

\newcommand{\dRSq}{{\overline{\delta R^2}}}
\newcommand{\dtSq}{{\overline{\delta \tau^2}}}
\DeclareMathOperator{\extr}{extr}
\DeclareMathOperator{\erfc}{erfc}
\DeclareMathOperator{\erf}{erf}

\begin{document}
\title{A collective phase in resource competition in a highly diverse ecosystem}
\author{Mikhail Tikhonov}
    \affiliation{School of Engineering and Applied Sciences}
    \affiliation{Kavli Institute for Bionano Science and Technology, Harvard University, Cambridge, MA 02138, USA}
\author{Remi Monasson}
    \affiliation{Laboratoire de Physique Th\'eorique de l'\'Ecole Normale Sup\'erieure -- UMR 8549, CNRS and PSL Research, Sorbonne Universit\'e UPMC, 24 rue Lhomond, 75005 Paris, France}

\begin{abstract}
Organisms shape their own environment, which in turn affects their survival. This feedback becomes especially important for communities containing a large number of species; however, few existing approaches allow studying this regime, except in simulations. Here, we use methods of statistical physics to analytically solve a classic ecological model of resource competition introduced by MacArthur in 1969. We show that the non-intuitive phenomenology of highly diverse ecosystems includes a phase where the environment constructed by the community becomes fully decoupled from the outside world.
\end{abstract}
\maketitle
Understanding the diversity of life forms on our planet is an age-old question. Recent technological advances uncovered that most habitats harbor hundreds of coexisting ``species'' (most of which are microbial~\cite{Gill06,Caporaso11,Lozupone12}), and the problem of understanding such communities is currently at the forefront of medical and environmental sciences~\cite{HMP,EMP,Beiko15}. One of the key obstacles arises from the fact that ecological and evolutionary time scales are generally not separable, giving rise to a coupled ``eco-evolutionary dynamics''~\cite{Fussmann07,Pelletier09,Henson15}. The fitness of an organism depends on its environment, but this environment is not fixed: it includes all other organisms in the community, is shaped by their activity and changes on an ecological time scale. Understanding this feedback has long been recognized as an important question of community ecology~\cite{Schoener11}.

A convenient example of such ecological feedback appears in models of resource competition~\cite{Grover97}. The survival of an organism is determined by the availability of resources in its immediate environment. In quantitative theories of evolution (population genetics), we typically think of this environment as being fixed externally, but in an ecological setting an experimentalist can only set the conditions faced by the community as a whole, e.g.\ the overall influx of resources. The immediate environment of an individual is affected by the activity of all other organisms and is not under our direct control. For example, consider increasing the overall influx of maltose (a sugar) to a multi-species bacterial culture. This could lead to an increase of maltose in the medium, opening the community to invasion by a species that grows well on this sugar. Alternatively, this could enable existing maltose-consuming species to expand in population, driving maltose availability back to the same level, or perhaps even depleting it further. The relation between the resources supplied to the community and the immediate environment seen by individual organisms is non-trivial. Our control extends on the former, but organism survival and therefore community structure are determined by the latter.

The mechanisms by which organisms shape their environment (niche construction theory~\cite{ScottPhillips14}) have been the subject of much research, both at equilibrium (e.g.\ resource competition models~\cite{Grover97}) and out of equilibrium (e.g.\ in the study of ecological successions~\cite{McCook94}). Perhaps the most progress was achieved in the problem of resource competition in a well-mixed community at equilibrium, introduced 50 years ago by MacArthur~\cite{MacArthur}. However, the geometric approach developed by Tilman in his classic work~\cite{Tilman82} allowed him to analyze only the cases with $N=1$ and $N=2$ resources. It is not clear to what extent the intuition derived from low-dimensional models applies to the high-dimensional case. Recently, a simulation-based study of a modestly larger number of resources ($N=10$) exhibited a surprising effect whereby a community interacting with another community would exhibit an effective ``cohesion'' even in the absence of any cooperative interactions between its members, purely as a consequence of environmental feedback~\cite{CWC}. The number of metabolites at play in a complex microbial community in nature is even larger, of order $N\simeq 100$~\cite{Fischbach07,Fischbach11}. It is an intriguing possibility that the phenomenology of high-diversity communities could contain qualitatively novel, non-intuitive regimes. However, few existing approaches allow studying niche construction or eco-evolutionary dynamics for a large number of interacting species, except in simulations.

In this work, we show that MacArthur's classic model of resource competition can be formulated as a problem of statistical physics of a disordered system, and solved analytically in the limit of large $N$. We observe a phase transition between two qualitatively distinct regimes. In one regime, changes of external conditions propagate to the immediate environment experienced by organisms, as expected. However, in the other regime, the immediate environment of individuals organisms becomes a collective property of the community, unaffected by the outside world. This regime, which only arises at sufficient diversity, documents the emergence of a collective behavior as a consequence of large dimensionality.

In defining our model, we follow Ref.~\cite{CWC}, but allow for more generality.
Consider a multi-species community in a well-mixed habitat where a single limiting element $\XX$ exists in $N$ forms (``resources'' $i\in\{1\dots N\}$). For example, this could be carbon-limited growth of bacteria in a medium supplied with $N$ sugars. Let $n_\mu$ denote the population size of species $\mu\in\{1\dots \mathcal S\}$. Briefly, the availability $h_i$ of each resource $i$ in the immediate environment of individuals will determine the dynamics of $n_\mu$. The changes in species abundance will translate into changes in the total demand for resources, denoted $T_i$. This total demand, in turn, will determine the resource availability $h_i$. This feedback loop is the focus of our analysis.

A species is characterized by its requirement $\chi_\mu$ for the limiting element $\XX$, and the ``metabolic strategy'' $\{\sigma_{\mu i}\}$ it employs to try and meet this requirement. We think of $\sigma_{\mu i}$ as the investment of species $\mu$ into harvesting resource $i$ (e.g., the expression level of the corresponding metabolic pathway). Specifically, for given resource availability $\{h_i\}$, the population growth rate of species $\mu$ is determined by the \textit{resource surplus} $\Delta_\mu$ experienced by its individuals:
\begin{equation}\label{eq:surplus}
\frac{dn_\mu}{dt}\propto n_\mu\Delta_\mu \quad\text{with}\quad \Delta_\mu= \sum_i \sigma_{\mu i}\, h_i - \chi_\mu.
\end{equation}
The first term is the total harvest of $\XX$ from all sources, and the second is the requirement an individual must meet to survive.
The proportionality coefficient is not important, since we will only be concerned with the equilibrium state where $\frac{dn_\mu}{dt}=0$.

Species abundances $n_\mu$ determine the total resource demand $T_i\equiv \sum_{\mu} n_\mu \sigma_{\mu i}$. This demand shapes resource availability $h_i$. In the simplest model~\cite{CWC}, organisms could be sharing a fixed total influx of resource $R_i$: $h_i(T_i)=R_i/T_i$. In his original formulation, MacArthur considered a more complex scenario of dynamical resources with renewal rate $r_i$ and maximal availability $K_i$; this would correspond to setting
$
h_i(T_i)=K_i\left(1-\frac {T_i}{r_i}\right),
$
see eq.~(3) in Ref.~\cite{MacArthur}. In the interest of generality, here we will say only that the availability of resource $i$ is a decreasing function of this total demand: $h_i=H_i(T_i)$, and allow the functions $H_i(\cdot)$ to remain arbitrary, and possibly different for each resource.

\begin{figure}[b!]
\centering
\includegraphics[width=0.95\linewidth]{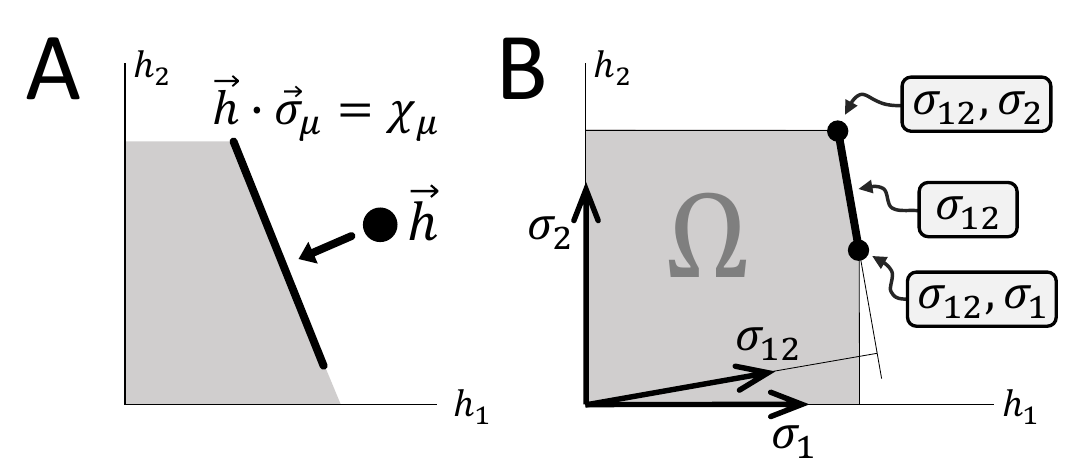}
\caption{The geometry of resource competition at $N=2$. \textbf{A:}~If resource availability $\vec h$ lies above the line $\vec h\cdot\vsigma_\mu=\chi_\mu$, the species $\mu$ will multiply, depleting resources (arrow). \textbf{B:}~Competition between $\mathcal S=3$ species; metabolic strategies indicated by arrows (two specialists and one mixed strategy). The equilibrium $\vec h$ is always located at the boundary (highlighted) of the ``unsustainable region'' $\Omega$; one or two species may coexist.\label{fig:geom}}
\end{figure}

This model admits a convenient geometric formulation, where we can think of the metabolic strategies $\{\sigma_{\mu i}\}$ as $\mathcal S$ vectors in the $N$-dimensional space of resource availability. Each hyperplane $\vec h\cdot\vec \sigma_\mu=\chi_\mu$ separates this space into two regions (Fig.~\ref{fig:geom}A). Above this hyperplane, a positive resource surplus allows species $\mu$ to multiply. Below this hyperplane (shaded), resources are insufficient to support species $\mu$. The intersection of such regions over all competing strategies $\{\vsigma_\mu, \chi_\mu\}$ defines the ``unsustainable region'' $\Omega$:
$$
\Omega = \bigcap_{\mu=1}^{\mathcal S}\; \{\vec h \;|\; \vec h \cdot \vsigma_\mu<\chi_\mu\}
$$
If resource availability $\vec h$ is inside $\Omega$, no species can harvest enough resources to sustain its population. Outside $\Omega$, at least one species can increase its abundance. Therefore, the equilibrium state can only be located at the boundary of $\Omega$, which we denote $\partial\Omega$. The dynamics~\eqref{eq:surplus} possesses a Lyapunov function, which is convex and bounded from above, similar to the classic model of MacArthur of which this is a generalization (see SI). As a result, the equilibrium state always exists, is unique and stable, and can be found by solving a convex optimization problem over the region $\partial \Omega$. At this equilibrium, each species is either extinct and cannot invade ($n_\mu=0$, $\Delta_\mu<0$), or is present and its resource balance is met ($n_\mu>0$, $\Delta_\mu=0$).

Fig.~\ref{fig:geom}B shows an example at $N=2$. Here, a community of two specialists $\vec\sigma_1=\{1,0\}$ and $\vec\sigma_2=\{0,1\}$, both with cost $\chi_0$, is exposed to a mixed strategy $\vsigma_{12}=\{x,1-x\}$
with a cost slightly below $\chi_0$. The species $\vsigma_{12}$ will be able to invade, and depending on resource supply, may coexist with one of the specialists (but not both). The equilibrium will harbor one or two species, corresponding to the equilibrium $\vec h$ being located either at an edge or at a vertex of $\partial\Omega$.

The resource depletion rules $H_i(\cdot)$ describe the external conditions: 
how much of each resource is supplied to the community as a whole. In contrast, $\vec h$ describes the availability of resources in the immediate environment of individuals, which ultimately dictates which species survive. 
Any set of competing strategies $\{\vec\sigma_\mu, \chi_\mu\}$ defines a unique community equilibrium, and so implements a mapping from external conditions into the actual environment $\vec h$. Our aim is to characterize the properties of this mapping.

The geometric intuition described above was first developed by Tilman~\cite{Tilman82}, who exhaustively analyzed the cases $N=1$ and $N=2$. In higher dimensions, however, the enumeration of co-existence regimes for a given set of strategies, like in Fig.~\ref{fig:geom}B, quickly becomes a combinatorially difficult problem. In this work, we therefore adopt the statistical physics approach, and characterize the expected properties of a \textit{typical} community, when the competing strategies are drawn out of some ensemble.

Specifically, for each species $\mu$, we first pick its strategy as a random binary vector, where each component $\sigma^i_\mu$ is 1 with probability $p$, and 0 otherwise. The parameter $p$ allows us to specify the location of a typical competitor on the specialist-generalist axis. We then draw a random cost $\chi_\mu = \big(\sum_i\sigma_{\mu i}\big) + \epsilon x_\mu$, where $\epsilon$ is a parameter (the cost scatter, assumed small), and $x_\mu$ is a Gaussian random variable of zero mean and unit variance. We set the total number of species to $\mathcal S\equiv \alpha N$.

The key simplification that makes the problem tractable analytically is the independence of $\vsigma_\mu$ and $x_\mu$: the strategy and its cost are effectively uncorrelated. This assumption is strong, but far from unreasonable. 
The species competing for the same resources in real communities differ in evolutionary history, lifestyle, and physiology. Modeling the cumulative effect of these differences as a random contribution to the species' likelihood to succeed is arguably a better null model than claiming that the single factor we explicitly consider (the species' metabolic preference) plays the dominant role in determining its intrinsic performance.


Note that setting $h_i=1$ satisfies the resource balance of all species within a quantity of order $\epsilon$, so this cost model ensures that neither specialists not generalists have an obvious advantage~\cite{CWC}. To characterize the fluctuations of resource availability $1-h_i$, we introduce:
$$
m=\sum_i (1-h_i),
\qquad
q=\sum_i (1-h_i)^2.
$$
The resource surplus of a typical species is given by:
$$
\langle\Delta_\mu \rangle = \left\langle\sum_i h_i\sigma_{\mu i}-\Big[\sum_i \sigma_{\mu i}+\epsilon x_\mu\Big] \right\rangle = -pm
$$
(the angular brackets denote the mean over $\mu$). Negative for most species, $\Delta_\mu$ should hit zero for the lucky outliers who survive. We find that the spread of resource surplus values is given by $\psi\equiv\sqrt{p(1-p)q+\epsilon^2}$ (see SI). Intuitively, this is because species differ in cost (variance $\epsilon^2$), and their strategy ($\{\sigma_{\mu i}\}$ with variance $p(1-p)$) picks out resources with different availability (variance $q$). For this reason, rather than using $q$ and $m$ directly, for our order parameters we choose $\psi$ and the ratio $\lambda\equiv \frac{pm}\psi$.

Each particular set of competitors constitutes ``frozen disorder'', and the properties of a typical community can be computed using methods of statistical physics of disordered systems~\cite{StatMechOfLearning}, as detailed in the SI.
For simplicity, all the results will be quoted for the simplest supply model $H_i(T_i)=\frac {R_i}{T_i}$ where each resource is characterized by a single parameter: its total supply $R_i$ (see SI for the general case).
Our calculation yields explicit equations for the order parameters $\psi$ and $\lambda$ at equilibrium, in the thermodynamic limit $N,\mathcal S\rightarrow\infty$ at $\alpha$ held constant:
\begin{equation*}\label{eq:syst}
\begin{aligned}
&\frac{1-\alpha I(\lambda)}{1-\alpha E(\lambda)} = 1+(1-p)\frac{\lambda}{\psi}\\
&\psi^2\big[1-\alpha I(\lambda)\big]=\epsilon^2+p(1-p)\dRSq\big[1-\alpha E(\lambda)\big]^2
\end{aligned}
\end{equation*}
Here $\dRSq$ is the variance of resource supply $R_i$, and
$I(\lambda)\equiv\int_0^\infty\!y^2e^{-\frac{(y+\lambda)^2}2}\frac{dy}{\sqrt{2\pi}}$ and $E(\lambda)\equiv\int_\lambda^\infty\!e^{-\frac{y^2}2}\frac{dy}{\sqrt{2\pi}}$ are auxiliary functions that can be expressed in terms of the error function $\erf$.

\begin{figure}[t!]
\centering
\includegraphics[width=\linewidth]{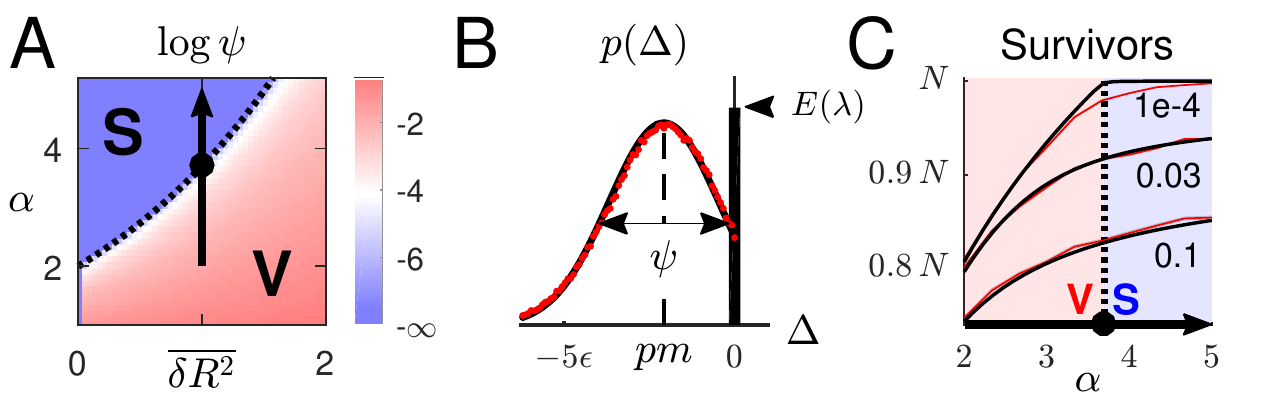}
\caption{\textbf{A.} The phase transition at $\epsilon\rightarrow 0$. In the $S$-phase, above a critical $\alpha$ (dotted line), the fluctuations of internal resource availability $\psi$ vanish, shown here on log scale to highlight the transition. \textbf{B.} The distribution of resource surplus at equilibrium. Black, the theoretical prediction; red, simulation data accumulated over 500 realizations at $N=50$ and is shown for extinct species only (see SI for details). \textbf{C.} The number of surviving species at equilibrium as a function of $\alpha$ at $\dRSq=1$ (\textit{cf.} the arrow in panel A). Theoretical prediction (black); mean over 500 simulations at $N=50$ (red); the deviation at $\epsilon=10^{-4}$ is an effect of small $N$. Standard error of the mean is too small to be visible. Dotted line at critical $\alpha$; shading labels the two phases.\label{fig:phases}}
\end{figure}

To study these equations, consider first the limit $\epsilon\rightarrow0$. In this limit, the parameter space separates into two phases (Fig.~\ref{fig:phases}A). One of these corresponds to the solution $\psi=1-\alpha E(\lambda)=0$ and will be called the S-phase; the other has $\psi\neq0$ and will be called the V-phase. The critical line (dotted line in Fig.~\ref{fig:phases}A) is described by:
\begin{equation*}
\dRSq_{\text{crit}} = \frac {1-p}p\frac{\lambda^2 }{1-\alpha_{\text{crit}}\,I(\lambda)}, \text{ where }\lambda=\frac{1}{E^{-1}(\alpha_{\text{crit}})}
\end{equation*}
For $\dRSq=0$ the transition occurs at $\alpha_{\text{crit}}=2$, consistent with the perceptron phase transition~\cite{StatMechOfLearning,Gardner88}.

To understand the physical meaning of these phases, consider first a community consisting of $N$ perfect specialists with costs $\chi_\mu\equiv1$. This community constitutes an example of the S-phase, where the immediate environment of individuals is fully ``\textbf{s}hielded'' from external conditions: faced with an uneven resource supply, species' abundance will adjust to drive resource availability to $h_i=1$ for all $i$, restoring symmetry. In general, a restricted set of species (small $\alpha$) or a strongly heterogeneous resource supply (large $\dRSq$) will prevent the community from exactly matching demand to the uneven supply, and the externally imposed asymmetry between resources will propagate into the organisms' actual environment $\vec h$ (the V-phase, ``\textbf{v}ulnerable'' to external perturbations). However, as the community is exposed to new species ($\alpha$ is increased above the critical value; the arrow in Fig.~\ref{fig:phases}A), the community transitions into the shielded phase where the environment $\vec h$ is fully symmetric ($m=q=0$) and insensitive to external conditions.

To confirm this interpretation, consider the number of coexisting species at equilibrium. As we have seen, geometrically, this number is the co-dimension ($N$ minus the dimension) of the region of $\partial\Omega$ where the equilibrium is located. Remarkably, this elusive quantity can also be computed analytically. Specifically, one can compute the distribution of the resource surplus $\Delta$ of all $\alpha N$ species at equilibrium (Fig.~\ref{fig:phases}B; see SI):
$$
p(\Delta)=\frac{1}{\sqrt{2\pi\psi^2}}e^{-\frac{(\Delta+\lambda\psi)^2}{2\psi^2}}\cdot\theta(-\Delta)+E(\lambda)\delta(\Delta),
$$
Here $\theta$ is the Heaviside function constraining $\Delta$ to be negative. The delta-shaped peak at $\Delta=0$ represents the fraction of species whose resource demand is met. The number of survivors is therefore $\alpha N\, E(\lambda)$, in excellent agreement with simulations (Fig.~\ref{fig:phases}C). The S-phase where $\alpha E(\lambda)=1$ therefore harbors a complete set of exactly $N$ species. 
If the perturbation of external conditions is small, no species will go extinct. Since the vectors $h_i$ and $\chi_\mu$ ($\mu$ running over $N$ surviving species) are related by a full-rank matrix $\sigma_{\mu i}$, this means that the resource availability at the new equilibrium will remain exactly the same, confirming our interpretation of this ``shielded'' phase.

\begin{figure}[t!]
\centering
\includegraphics[width=\linewidth]{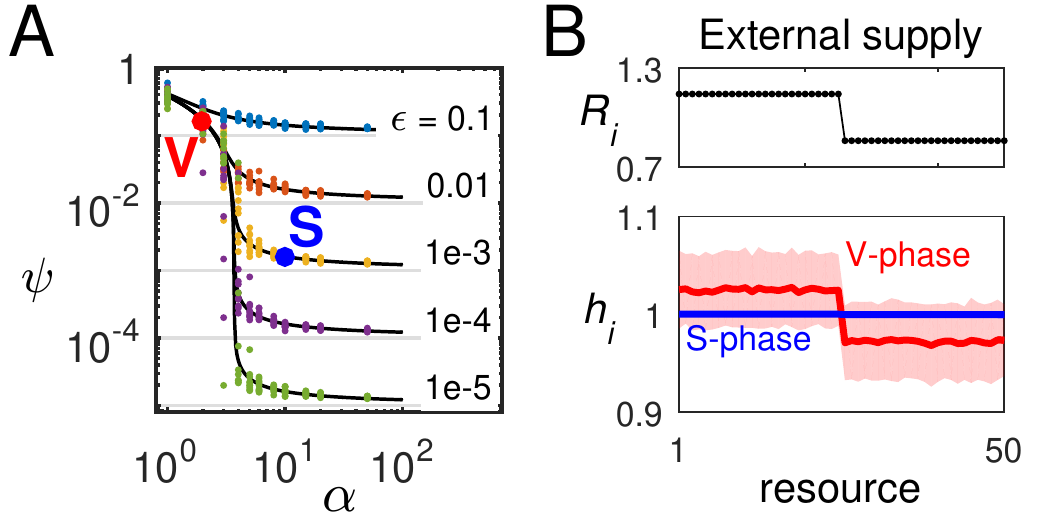}
\caption{
\textbf{A.} At finite $\epsilon$, the phase transition is replaced by a crossover. Theoretical curves are overlaid with simulation datapoints for a range of $\alpha$ (10 instances each). At large $\alpha$, we observe $\psi\rightarrow\epsilon$, confirming that the fluctuations of $h_i$ become negligible.
\textbf{B.} The qualitative distinction between phases persists at finite $\epsilon$. Here, simulation results are shown for $\epsilon=10^{-3}$. A community faces a bimodal supply of $N=50$ resources (upper panel). Lower panel shows the equilibrium availability of resources $h_i$ (mean $\pm$ 1 standard deviation over 500 instances), for two values of $\alpha$ corresponding to different phases (highlighted in panel A). In the ``shielded'' S-phase, the asymmetry of the external supply does not affect resource availability $h_i$.\label{fig:cross}}
\end{figure}

For a non-zero $\epsilon$, the strict phase transition is replaced by a crossover (Fig.~\ref{fig:cross}A). The role of $\epsilon$ in our model is to measure how strongly a species' fate is influenced by intrinsic, rather than environment-dependent (ecological) factors~\cite{CWC}. For large $\epsilon$, community structure is no longer shaped by interactions between community members, but becomes dominated by species who outperform others in all circumstances, and the environmental feedback studied here becomes irrelevant. For small $\epsilon$, however, the distinct features of the ``shielded'' and ``vulnerable'' phases remain clearly recognizable: the fluctuations of resource availability are, respectively, of order $\epsilon$ or much larger than $\epsilon$ (Fig.~\ref{fig:cross}B).

This result has intriguing implications. Consider a community facing the strongly uneven resource supply shown in Fig.~\ref{fig:cross}B (top panel). Define a species' individual performance as its growth rate when placed in this environment, with no other organisms present. One might expect this performance metric to be predictive of species' survival in a community setting: surely, increasing the supply of maltose to a community should favor organisms that grow well on maltose. In the more intuitive V-phase, this expectation is indeed correct. However, in the S-phase the internal environment becomes a collective property governed by the statistical properties of the species' pool, rather than by the external conditions (Fig.~\ref{fig:cross}B; bottom panel). As a result, the performance measured in external conditions becomes irrelevant: it no longer predicts whether a species will survive (Fig.~\ref{fig:SXXphases}).

In ecological terms, the model considered here was purely competitive: increasing the abundance of any species reduces the growth rates of everyone else, i.e.\ there are no ``cooperative interactions''. Nevertheless, we have shown that at high dimension, the parameter space of this classic resource competition model contains a strongly collective regime.

These conclusions were drawn in the context of a particular, highly simplified model. In particular, our analysis ignored spatial structure, assumed deterministic dynamics, and considered the equilibrium states only. It is clear that natural communities are never in steady state, and stochasticity and spatial structure are tremendously important in most contexts. Nevertheless, the goal of this work was to explore specifically the feedback of organisms onto their environment and identify the implications of large dimensionality. For this purpose, the simplified model adopted here provides a convenient starting point, and highlights the promise of applying statistical physics to gain analytical insight into the non-intuitive phenomenology of large-dimensional networks~\cite{DeMartino10} and highly diverse ecosystems.

We thank Michael P. Brenner, Carl P. Goodrich, Alpha Lee, Emily Zakem and David Zwicker for helpful discussions, the Harvard Center of Mathematical Sciences and Applications, and the Simons Foundation. This work was completed at the Aspen Center for Physics, supported by National Science Foundation grant PHY-1066293.

\onecolumngrid
 \cleardoublepage
 \section*{Supplementary material}
 \setcounter{figure}{0}
 \setcounter{equation}{0}
 \setcounter{section}{0}
 \renewcommand{\thesection}{S\arabic{section}}
 \renewcommand{\thesubsection}{S\arabic{section}.\arabic{subsection}}
 \renewcommand{\theequation}{S\arabic{equation}}
 \renewcommand{\thefigure}{S\arabic{figure}}
 \renewcommand{\thetable}{S\arabic{table}}
\section{The Lyapunov function $F$}\label{sec:Lyapunov}
Recall that the dynamics of our model are given by
$$
\frac{dn_\mu}{dt}=b_\mu n_\mu\Delta_\mu,
$$
where $\Delta_\mu$ is the resource surplus $\Delta_\mu = \sum_i \sigma_\mu^i H_i(T_i) - \chi_\mu$. This section will show that this dynamics possess a Lyapunov function:
\begin{equation}\label{eq:F}
F(\{n_\mu\})=\sum_i \hat H_i(T_i) - \sum_\mu n_\mu\chi_\mu.
\end{equation}
In other words, $F$ increases on any trajectory of the dynamics above. In addition, we will show that $F$ is convex and bounded from above.

\subsection*{Proposition 1: $F$ increases on any trajectory}
We first note that the derivative of $F$ with respect to a species' abundance $n_\mu$ is precisely the resource surplus $\Delta_\mu$:
$$
\frac{\partial F}{\partial n_\mu}=\sum_i H_i(T_i)\frac{\partial T_i}{\partial n_\mu}-\chi_\mu=\Delta_\mu,
$$
Therefore, $F$ is indeed a Lyapunov function:
$$
\frac{dF}{dt}=\sum_\mu\frac{\partial F}{\partial n_\mu}\frac{dn_\mu}{dt}=\sum_\mu b_\mu n_\mu\Delta_\mu^2>0.
$$

\subsection*{Proposition 2: $F$ is bounded from above}
To see this, recall that $H_i(\cdot)$ was required to be a decreasing function of its argument; moreover, to forbid unbounded growth of any species, we required that for large enough demand $T$, the resource availability $H_i(T)$ should go to zero. It follows that its integral $\hat H_i(x)\equiv\int^x\!H(T)\,dT$ grows sub-linearly; in other words, for any $\lambda>0$ we have $H_i(x)<\lambda x$ if $x$ is large enough. We conclude that $F(\vec n)$ goes to $-\infty$ as the norm of the abundance vector increases (this precisely corresponds to forbidding infinite population growth). A continuous function defined on the positive quadrant $\{n_\mu\ge0\}$ and going to $-\infty$ at the boundary of this region is bounded from above, as claimed.

\subsection*{Proposition 3: $F$ is convex}
To see this, note that for any function $f(\vec n)$, the following two operations leave its convexity invariant ($M$ is an arbitrary matrix):
\begin{enumerate}
\item adding a linear function of its arguments:
    $f(\vec n) \mapsto g(\vec n) = f(\vec n)+M\vec n;$
\item performing a linear transformation of its arguments:
    $f(\vec n) \mapsto h(\vec n) = f(M\vec n).$
\end{enumerate}
Given these observations, convexity of $F$ directly follows from the convexity of $\hat H_i(x)$ (which is an integral of a decreasing function).
\section{Locating the community equilibrium: the geometric intuition}\label{sec:geomIntuition}
The main text shows that the equilibrium of community dynamics is always located at the boundary of the ``unsustainable region'' $\Omega$ defined in the text. Which boundary point is selected? Here we present an intuitive geometric argument, which will be formalized in the following section.

\begin{figure}[h!]
\centering
\includegraphics[width=0.5\linewidth]{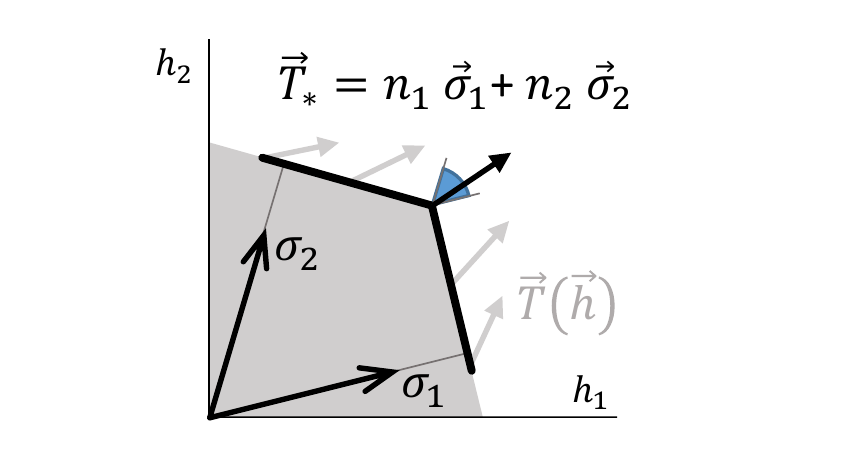}
\caption{The geometric intuition behind the selection of the equilibrium point: at equilibrium, the vector of total demand $\vec T_*$ must be pointing ``strictly outward'' relative to the unsustainable region $\Omega$. Here, at a two-species equilibrium of $\vsigma_1$ and $\vsigma_2$, $\vec T_*$ must lie within the sector highlighted in blue. The equilibrium point can therefore be found by following the vector field $\vec T(\vec h)$ (in gray).}\label{fig:SXXgeom}
\end{figure}

Let $h_*$ be the resource availability at community equilibrium. For concreteness, consider the case $N=2$, and assume the equilibrium state harbors two species $\{\vsigma_1,\chi_1\}$ and $\{\vsigma_2,\chi_2\}$, so that the point $\vec h_*$ is the intersection of lines $\vec h\cdot \vsigma_1=\chi_1$ and $\vec h\cdot \vsigma_2=\chi_2$. Consider now the vector of total demand $\vec T_*$ at this equilibrium. By definition, it is a linear combination of the two strategy vectors: $\vec T_*=n_1\vsigma_1+n_2\vsigma_2$. Importantly, the coefficients here must be \textit{positive}. We conclude that at equilibrium, the vector $\vec T_*$ must point ``strictly outward'' relative to the region $\Omega$, as in Fig.~\ref{fig:SXXgeom}.

This property is sufficient to uniquely determine the equilibrium point. Indeed, consider a vector field $\vec T(\vec h)$, where to each point of the resource availability space $\vec h_0$ we associate the vector of total demand $\vec T_0$ that corresponds to such resource depletion, i.e.\ such that $H(\vec T_0) = \vec h_0$. The intuitive argument above suggests that the equilibrium of community dynamics can be found by following this vector field. And indeed, this vector field is a gradient of a certain function, and therefore locating the equilibrium corresponds to maximizing this function. This is formally proven in the following section.

We stress that the vector field $\vec T(\vec h)$ does not describe the dynamics itself; it is merely a tool to find the equilibrium point. The trajectories of the system in the harvest space are \textit{not} integral lines of this vector field.

\section{Locating the community equilibrium: the formal proof}\label{sec:Legendre}
\textbf{Proposition 1:} \textit{There exists a function $\tilde F$ defined on the harvest plane, such that its gradient at any point $\vec h$ is the demand vector $\vec T$ that corresponds to this resource availability vector:}
$$
\frac{\partial \tilde F}{\partial h_i} = T_i\qquad\Leftrightarrow\qquad H_i(T_i)=h_i.
$$

\textbf{Proof:} Consider $F=\sum_i \hat H_i(T_i)$. This function has the property that $\frac{\partial F}{\partial T_i} = H_i(T_i)$. The function $\tilde F$ can be explicitly constructed as the Legendre transform of $F$:
$$
\tilde F(\vec h) = \left.\left[\vec h\cdot\vec T-F(T)\right]\right|_{\vec T=\vec T^*}
$$
where $\vec T_*$ is defined by the condition $H_i(T^*_i)=h_i$. It is easy to check that this function satisfies the desired requirement. Indeed, for each component $i$ (and omitting this index for simplicity):
$$
\frac{\partial\tilde F}{\partial h}=T^*+h\frac{\partial T^*}{\partial h}-
\frac{\partial T^*}{\partial h}\left.\frac{\partial\hat H}{\partial T}\right|_{T^*}=T^*.
$$

Consider now an equilibrium community $\CC$ with the total demand $\vec T^*$, and the resource availability vector $\vec h_*$. We already know that $\vec h_*$ lies at the boundary of the unsustainable region $\Omega$ (see main text). To determine exactly which boundary point is selected, we make the following observation:

\textbf{Proposition 2:} \textit{For any other vector $\vec h_1\in\Omega$, we have}
\begin{equation}\label{eq:projection}
(\vec h_1-\vec h_*)\cdot\vec T\le0.
\end{equation}

\textbf{Proof:} Since $\vec h_*$ is the equilibrium state, we can write:
$$
\forall \mu:\qquad n_\mu\left[\vsigma_\mu\cdot \vec h_*-\chi_\mu\right]=0.
$$
As for $\vec h_1$, it lies in the unsustainable region $\Omega$, and therefore:
$$
\forall \mu:\qquad n_\mu\left[\vsigma_\mu\cdot \vec h_1-\chi_\mu\right]\le 0.
$$
Subtracting the former from the latter, and summing over $\mu$, we conclude:
$$
\sum_\mu n_\mu \vsigma_\mu(\vec h_1-\vec h_*)\le 0 \qquad \Rightarrow \qquad \vec T\cdot (\vec h_1-\vec h_*)\le 0  \text{ as claimed. $\square$}
$$
This means that the equilibrium point is such that the value of $\tilde F$ cannot be further increased: any movement within the unsustainable region goes against the gradient field of $\tilde F$.

\textbf{Corollary:}
\textit{The equilibrium community state corresponds to the maximum of $\tilde F$ over the unsustainable region $\Omega$.}

It is worth contrasting our approach to other situations where community-level objective functions may appear, e.g. community-level flux balance analysis, or similar approaches. In certain contexts, an optimization-based framework is simply postulated, and serves as an exploratory tool to investigate the possible regimes of network performance: for instance, the total metabolic output of a consortium might be taken as a ``plausible'' global objective function for the community to optimize. Here, we stress that no community-level objectives are postulated; the fact that the ecological dynamics in this model take the form of a global optimization problem is a ``lucky'' consequence of explicitly specified dynamics of (purely ``selfish'') individual species. This special feature makes MacArthur's resource competition model an especially convenient starting point for investigating the consequences of high dimensionality in an ecological context.

\section{The algebra of ``passing into the harvest space''}
Above, we have shown that locating the equilibrium of our ecological dynamics is in fact a convex optimization problem in the $P$-dimensional space of species abundances. We then described how this optimization problem can be formulated directly in the $N$-dimensional space of ``harvests'' $h_i$. To build intuition, it is instructive to consider the following algebraic argument showing how the two optimization problems are mapped into each other.

In order to locate the maximum of the Lyapunov function $F$, we investigate the large-$\beta$ limit of the partition function $Z(\beta)$:
$$
\max F = \lim_{\beta\rightarrow\infty}\left( \frac {\log Z}\beta\right),\text{ where }Z(\beta)=\int_0^\infty\! e^{\beta F}\prod_\mu dn_\mu
$$
To compute $Z$, we first introduce $T_i$ as convenient auxiliary variables. This allows integrating over $n_\mu$:
$$
\begin{aligned}
Z&=\int_0^\infty\!\prod_\mu dn_\mu\int_0^\infty\!dT_i\,\delta\Big(T_i-\sum_\mu n_\mu\sigma^i_\mu\Big) e^{\beta F(n_\mu)}\\
 &=\int_0^\infty\!\prod_\mu dn_\mu\int_0^\infty\!dT_i\left[\int\!\!\frac{d\theta_i}{2\pi/\beta}\,e^{-i\beta\theta_i\big(T_i-\sum_\mu n_\mu\sigma^i_\mu\big)}\right] e^{\beta \big(\sum_i \hat H_i(T_i) - \sum_\mu n_\mu\chi_\mu\big)}\\
 &=\int_0^\infty\!\!dT_i\int\!\!\frac{d\theta_i}{2\pi/\beta}\,
 e^{\beta\sum_i \left[\hat H_i(T_i)-i\theta_i T_i\right]} \prod_\mu \int_0^\infty\!\!\! dn_\mu\,
 e^{-\beta n_\mu\left[\chi_\mu-i\sum_i\theta_i\sigma^i_\mu\right]}\\
 &=\int_0^\infty\!\!dT_i\int\!\!\frac{d\theta_i}{2\pi/\beta}\,
 e^{\beta\sum_i \left[\hat H_i(T_i)-i\theta_i T_i\right]}\prod_\mu \frac{1/\beta}{\chi_\mu-i\sum_i\theta_i\sigma^i_\mu}\\
\end{aligned}
$$
We now focus on the integral over $T_i$. For large $\beta$, it can be computed using saddle-point method. Denoting $i\theta_i\equiv h_i$, we find that the saddle-point $T_i^*$ is defined by the condition:
$$
H_i(T_i^*)=h_i,
$$
which justifies our suggestive notation (we recognize $h_i$ as the substrate availability at equilibrium demand $T_i^*$). This condition implicitly defines $T_i^*$ as a function of $h_i$, so all that remains is the $N$-dimensional integral over $\vec h$:
\begin{equation}\label{eq:inHarvestSpaceHalfway}
Z=\const\times\int_{-i\infty}^{i\infty}\! d\vec h\, \exp\left[-\beta\tilde F\big(\vec h\big)\right]\prod_\mu \frac {1/\beta}{\chi_\mu-\vec h\cdot \vsigma_\mu}.
\end{equation}
The $\tilde F$ in the exponent is precisely the Legendre transform of $\hat H_i$:
$$
\tilde F\equiv \sum_i \tilde F_i = \sum_i\left[h_i T_i-\hat H_i(T_i)\right]_{\text{at } T_i=T_i^*}.
$$
For large $\beta$, this is again a saddle-point integral. We are starting to recognize the problem of extremizing $\tilde F$; however, here it is computed for purely imaginary arguments, and so a few more steps are needed. The integration contours cannot simply be rotated onto the real axes, since the integrand has a complicated pole structure. Instead, we can convert the integration contours into piecewise-linear shapes, two of which are purely imaginary, and one is purely real: $-i\infty \rightarrow 0 \rightarrow x \rightarrow x+i\infty$, with $x\in\mathbb R$. The deformation of the integration contour is allowed only as long as the poles are not crossed, and the integrand has a pole whenever $\Delta_\mu=0$ (we note that the denominator in~\eqref{eq:inHarvestSpace} is $(-\Delta_\mu)$, the negative resource surplus of species $\mu$). Thus in our $N$-dimensional integral, the shifting of each contour will depend on the exact values of all other variables. Thankfully, the integrand can have an extremum only if all $h_i$ are real, and whenever $N-1$ variables $h_i$ take real values, the remaining one can vary (on its real-valued portion of the contour) from $0$ to the highest value that can be reached without crossing any of the hyperplanes $\Delta_\mu=0$. The region delimited by these hyperplanes is precisely the ``unsustainable region'' $\Omega$ defined in the main text. We conclude that for the purposes of the saddle-point calculation, our integral becomes:
\begin{equation}\label{eq:inHarvestSpace}
Z=\const\times\int_\Omega\! d\vec h\, \exp\left[-\beta\tilde F\big(\vec h\big)\right]\prod_\mu \frac {1/\beta}{\chi_\mu-\vec h\cdot \vsigma_\mu}.
\end{equation}
The certain lack of rigour in our description of the transition from Eq.~\ref{eq:inHarvestSpaceHalfway} to Eq.~\ref{eq:inHarvestSpace} will not be a problem. The purpose of this section is to build additional intuition about the algebraic structure of the problem, and analyzing the expression~\eqref{eq:inHarvestSpace} will prove instructive. However, the following sections will only use the fact that community equilibrium maximizes $\tilde F$, a result that was rigorously obtained in section titled ``Locating the community equilibrium: A formal proof''.

In expression~\eqref{eq:inHarvestSpace}, the exponential term $e^{-\beta\tilde F}$ dominates the integrand everywhere, except in the immediate vicinity of the region boundary where $\frac 1{\Delta_\mu}$ diverges. If $\beta$ is large, but finite, the extremum is achieved at a point $\vec h^*$ lying strictly inside the region $\Omega$, at a distance of order $1/\beta$ from the nearest bounding hyperplanes. In this ``finite temperature'' regime, all species $n_\mu$ have non-zero abundance: since $n_\mu$ enters into $Z$ as $e^{-\beta n_\mu\,|\Delta_\mu|}$, the observables $n_\mu$ follow an exponential distribution with mean $\langle n_\mu\rangle=\frac 1{\beta\,|\Delta_\mu|}$. In the zero-temperature limit ($\beta=\infty$), this expected abundance vanishes for all species except a select few, for which $\Delta_\mu$ is precisely zero. Thus, as described in the main text, the extremum $\vec h^*$ reaches the boundary of $\Omega$. At this value of harvests, a finite set of species have resource surplus of precisely zero, corresponding to finite-abundance survivors. The resource surplus of all other species is negative, and they go extinct at equilibrium.

To make this argument more precise, we note that at large, but finite $\beta$, Eq.~\eqref{eq:inHarvestSpace} gives us
$$
\log Z = \max_{\vec h\in\Omega} \left\{-\beta \tilde F(\vec h) -\sum_\mu \log |\Delta_\mu|\right\}.
$$
For a large $\beta$, the sum over $\mu$ is dominated by only a few terms, those corresponding to the closest hyperplanes for which $\Delta_\mu$ tends to zero. Denote their set $\mathcal S$ (for ``survivors''). The extremum condition:
\begin{equation}\label{eq:question}
\frac{\partial\tilde F}{\partial h_i}=-\sum_{\mu\in \mathcal S}\frac{1}{\beta|\Delta_\mu|}\frac{\partial|\Delta_\mu|}{\partial h_i} = \sum_{\mu\in \mathcal S} n_\mu \sigma^i_\mu = T_i.
\end{equation}
This of course makes perfect sense given the definition of $\tilde F_i$ as the Legendre transform of $\hat H_i$.

But if we are only interested in the identity of the species that survive at community equilibrium, it is wholly encoded in the location of the extremum $\vec h_*$ at $\beta=\infty$. Since this $\vec h_*$ is located at the boundary of $\Omega$, the shape of the repulsive potential of interaction with the hyperplanes $\Delta_\mu=0$ is irrelevant in this limit, and can be replaced by the Heaviside theta-function $\theta(-\Delta_\mu)$. In this limit, the problem reduces to computing the extremum of $\tilde F$ over the unsustainable region, as stated in the main text.

\section{Resource supply models}
\subsection{The model of MacArthur}
Different models of resource supply correspond to different expressions of the function $\tilde F$.  The renewable resource of MacArthur, with renewal rate $r$ and maximum resource availability $K$, is described by the following resource depletion rule (originally derived in Ref.~\cite{MacArthur}; see also the Supplementary section A in Ref.~\cite{CWC}):
$$
H(T)=K\left(1-\frac Tr\right)\qquad\Rightarrow\qquad T^*=r\left(1-\frac {h^*}K\right)
$$
Integrating $H(T)$, we find $\hat H(T)=KT-\frac {KT^2}{2r}$, and therefore
$$
hT^*-\hat H(T^*)= hr-\frac{r}{2K}h^2+\const
$$
We conclude that for the resource model of MacArthur:
$$
\tilde F(\{h_i\}) = \sum_i \left[r_ih_i -\frac{r_i}{2K_i}h_i^2\right]+\const
$$
\subsection{A constantly supplied resource}
The constant-supply model of Ref.~\cite{CWC} is a simpler model that postulates that a fixed amount of resource $R$ is evenly divided among all competitors: $H(T)=R/T$. In this model, we have $\hat H(T)=R\log T$ and $T^*=R/h^*$. Consequently:
$$
\tilde F = \sum_i h_i T^*_i-\hat H_i(T^*_i)=\sum_i R_i\log h_i+\const
$$
\subsection{A general model}
Consider the close-to-symmetric scenario, where the supply of all resources is similar. If the cost of all strategies is close to $\chi_0$, then the availability of resources at equilibrium will be close to $\chi_0$ as well. Linearizing around this point, a general resource supply model $H(T)$ can be characterized with two parameters. First, let $\tau$ be the value of demand at which resource availability hits $\chi_0$: by definition, $H(\tau)=\chi_0$. In the vicinity of this point, let $\gamma$ be the ``elasticity'' of supply, describing how quickly resource is depleted by a small increase in demand $\delta\tau\ll\tau$:
$$
H(\tau+\delta\tau)=\chi_0-\gamma\,\delta\tau.
$$
Let us compute the Legendre transform of $\hat H$ in this model. We have:
$$
H(T)=\chi_0-\gamma(T-\tau) \qquad\Rightarrow\qquad \hat H(T)=T(\chi_0+\gamma\tau)-\gamma\frac{T^2}2
$$
The demand that corresponds to a particular value of resource availability close to $\chi_0$:
$$
H(T^*)=h\qquad\Rightarrow\qquad T^* = \tau+\frac 1\gamma (\chi_0-h).
$$
In the vicinity of $h=\chi_0$ it is convenient to work with shifted variables: $h\equiv \chi_0-\frac {g}N$ (the $N$ in the denominator reminds that the deviation is small). After a little algebra we find:
$$
\tilde F_i(g_i) = -\frac12\gamma\tau_i^2 -\tau_i\frac {g_i}N - \frac 1{2\gamma_i}\left(\frac {g_i}N\right)^2,
$$
where the index $i$ reminds us that parameters $\tau_i$ and $\gamma_i$ could be different for different resources. Omitting the irrelevant global constant, we find the expression for $\tilde F$ in this general cost model:
\begin{equation}\label{eq:Fgeneral}
\tilde F_{\text{general}} = -\sum_i\left[\tau_i\frac {g_i}N + \frac 1{2\gamma_i}\left(\frac {g_i}N\right)^2\right].
\end{equation}

Of course, the two particular models we considered above reduce to this same form in the vicinity of $h_i\approx \chi_0$. Specifically, for MacArthur's model of renewable resource:
$$
\tilde F_{\text{MacArthur}}(\{g_i\}) = -\sum_i\left[r_i\left(1-\frac {\chi_0}{K_i}\right)\frac {g_i}N+\frac{r_i}{2K_i}\left(\frac{g_i}{N}\right)^2\right].
$$
Similarly, for the constant supply model:
$$
\tilde F_{\const}(\{g_i\}) = -\sum_i R_i\left[\frac {g_i}N+\frac12 \left(\frac{g_i}{N}\right)^2\right].
$$

Our calculation below will be for the general close-to-symmetric case where the supply of resources is similar:
\begin{equation}\label{eq:symmetricAnsatz}
\tau_i\equiv \bar \tau+\frac{\delta\tau_i}{\sqrt N} \qquad \gamma_i\equiv \bar \gamma+\frac{\delta\gamma_i}{\sqrt N}.
\end{equation}
Here $\sum_i\delta\tau_i=\sum_i\delta\gamma_i=0$ by definition, and as $N$ becomes large, $\delta\tau_i$ and $\delta\gamma_i$ remain of order 1. Note that the function $\tilde F$ can be rescaled by a constant positive factor, leaving the maximization problem unchanged (we seek the location of the maximum, not its magnitude). Without restricting generality, therefore, we can set $\bar\tau=1$.

\section{The cost model: an illustration}
\begin{figure}[h!]
\centering
\includegraphics[width=0.4\linewidth]{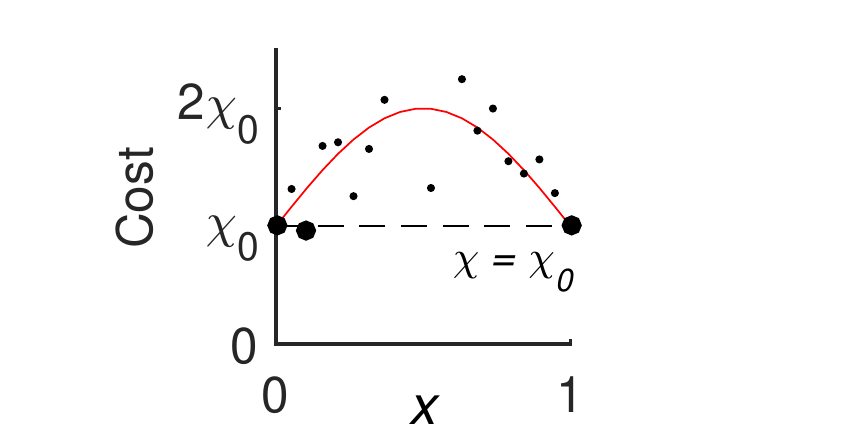}
\caption{\textbf{The cost model: an illustration.} Competition is restricted to a self-selected pool of low-cost outliers, so we only need to model this pool, where all strategies have similar costs. This illustration shows an example of a cost model at $N=2$ yielding the competition scenario depicted in Fig.~\ref{fig:geom}B of the main text (the cost of strategy $\{x,1-x\}$ (at $N=2$) is plotted as a function of $x$). Since metabolic diversification is penalized, all three competitive species (highlighted) are close to being specialists, but otherwise the details of the red curve have no effect on the coexistence problem.\label{fig:cost}}
\end{figure}

The main text made the argument that since competition is restricted to only a subset of species, the self-selected pool of low-cost outliers with similar costs $\chi_\mu$, the details of the cost model matter only inasmuch as they determine the properties of this subset. To illustrate this point, consider a scenario at $N=2$ where the cost of strategy $\{x,1-x\}$ is drawn out of a normal distribution with mean $\chi_0(1+\sin \pi x)$ and width $\frac12 \chi_0 \sin \pi x$. One such realization for 20 equally spaced values of $0\le x\le 1$ is shown in Fig.~\ref{fig:cost}. In this illustration, mixed strategies tend to be expensive; as a result, all three low-cost outliers (in bold) are close to specialists, and only these species are competitive (this is exactly the scenario depicted in Fig.~\ref{fig:geom}B). However, the exact details of the cost model (the precise shape of the solid red curve in Fig.~\ref{fig:cost}) are otherwise irrelevant for the coexistence problem.

\section{The replica-theoretic calculation}
This section demonstrates how the geometrical problem formulated above can be solved using methods of statistical physics, specifically an approach termed ``replica theory''. An attempt is made to present this computation in a detailed and self-contained way, i.e.\ not assuming familiarity with statistical physics of disordered systems. For a more comprehensive introduction to this powerful technique, we refer the reader to Ref.~\cite{StatMechOfLearning}.

\subsection{The basic idea}
We seek to compute:
\begin{equation}\label{eq:problem}
Z = \int_0^\infty \prod_i dh_i e^{\beta \tilde F}\prod_{\mu=1}^P \theta
\left(\chi^\mu-\vec h\cdot \vec\sigma^\mu\right)
\end{equation}

The pool of competitors is modeled as follows. For each species $\mu$, we pick $\vsigma_\mu$ as a random binary vector, where each component $\sigma^i_\mu$ is 1 with probability $p$, and 0 otherwise. We then draw a random cost $\chi_\mu = \sum_i\sigma^i_\mu + \epsilon x_\mu$, where $x_\mu$ is a Gaussian random variable of variance 1. 

The argument of the Heaviside $\theta$-function in Eq.~\eqref{eq:problem} is the negative resource surplus $-\Delta_\mu$. Under the cost model described above, we have
$$
\Delta_\mu=\vec h\cdot \vec\sigma^\mu-\chi^\mu=-\epsilon x_\mu - \sum_i\sigma^i_\mu(1-h_i).
$$
Change variables $h_i\equiv 1-\frac{g_i}N$ ($g_i$ runs from $-\infty$ to $N$), and introduce $\Delta_\mu$ as an explicit auxiliary integration variable:
$$
\begin{aligned}
Z &= \int_{-\infty}^N \prod_i \frac{dg_i}N e^{\,\beta \tilde F(\{g_i\})}\prod_{\mu=1}^P \int\! d\Delta_\mu\, \theta(-\Delta_\mu)\,
\delta\left(\Delta_\mu+\epsilon x_\mu + \frac 1N\sum_i g_i\sigma^i_\mu\right)\\
&= \int_{-\infty}^N \prod_i \frac{dg_i}N e^{\,\beta \tilde F(\{g_i\})}
\prod_{\mu=1}^P \int\! \frac{d\Delta_\mu\, d\hat\Delta_\mu}{2\pi}\,
\theta(-\Delta_\mu)\exp\left[i\sum_{\mu}\hat\Delta_\mu\left( \Delta_\mu+\epsilon x_\mu+\frac1{N}\sum_i g_i\sigma_\mu^i \right)\right]
\end{aligned}
$$
In this expression, $x^\mu$ and $\sigma_\mu^i$ are ``frozen disorder'': they are drawn randomly, but are then kept fixed, while other variables relax to their equilibrium values. Computing this integral for a particular realization of the disorder, even if it were possible to do so, would not be very informative. Instead, we are interested in the behavior of the ``typical'' realization of the system. This means that we are interested in the typical free energy $\langle F\rangle=\langle \log Z\rangle$ (angular brackets denote averaging over disorder). This quantity is hard to compute directly, because the average is outside of the logarithm. The opposite case, the logarithm of the average, would be very simple to compute; unfortunately, unlike the free energy $F$, the partition function $Z$ is dominated not by typical realizations of the disorder, but by extreme ones. The logarithm of the average would capture the behavior of the system in highly improbable extreme cases, which is of no use to us.

The gist of the ``replica trick'' is summarized in the following formula:
\begin{equation}\label{eq:trick}
\langle \log Z\rangle  = \lim_{n\rightarrow 0} \frac {\left\langle Z^n\right\rangle-1}n
\end{equation}
This trick makes it possible to formally derive an expression for $\langle \log Z\rangle$ by computing only expressions of type $Z^n$, which is the partition function of $n$ copies (``replicas'') of the system, and then formally sending $n$ to zero. The replicas are identical (have the same disorder), but independent (each has its own set of degrees of freedom). There are, of course, mathematical subtleties related to taking this limit, and we refer the reader to Ref.~\cite{StatMechOfLearning}. This reference also provide some intuition for the physical basis of the argument and the interpretation of the auxiliary variables that appear along the way. Here, the analytical result we derive will be validated by an excellent agreement with numerical simulations.

\subsection{Averaging over disorder}
Proceeding with our argument, we write the partition function of $n$ copies of our system. Each degree of freedom is copied $n$ times, labeled by the replica index $a$ running from 1 to $n$:
$$
\begin{aligned}
Z^n &= \int_{-\infty}^N \prod_{i,a} \frac{dg_i^a}N e^{\,\beta \sum_a\tilde F(\{g_i^a\})}
\prod_{\mu,a} \int\! \frac{d\Delta_\mu^a\, d\hat\Delta_\mu^a}{2\pi}\,
\theta(-\Delta_\mu^a)\exp\left[i\sum_{\mu,a}\hat\Delta_\mu^a\left( \Delta_\mu^a+\epsilon x_\mu+\frac1{N}\sum_i g_i^a\sigma_\mu^i \right)\right]
\end{aligned}
$$
We stress that all $n$ replicas have the same disorder: quantities $x_\mu$ and $\sigma_\mu^i$ have no index $a$. Thanks to our strategic choice of cost model, averaging $Z^n$ over these disorder variables is separable:
$$
\left\langle Z^n\right\rangle_{x_\mu,\vsigma_\mu} =
\int\prod_{i,a}\frac{d g_i^a}{N} e^{\beta\sum_{a}\tilde F(g_i^a)}
\prod_{\mu,a}\frac{d\Delta_\mu^a\, d\hat\Delta_\mu^a}{2\pi}\, \theta(-\Delta_\mu^a)\,e^{i\sum_{\mu,a}\hat\Delta_\mu^a \Delta_\mu^a}
\times
\underbrace{\prod_\mu\left\langle
e^{i\epsilon\sum_{a}\hat\Delta_\mu^a x_\mu}
\right\rangle_{x_\mu}}_{(1)}
\times
\underbrace{\prod_{i,\mu}\left\langle
e^{\frac{i}{N} \sum_{a}\hat\Delta_\mu^a g_i^a\sigma_\mu^i}
\right\rangle_{\sigma_\mu^i}}_{(2)}
$$
If $x$ is a Gaussian random variable of unit variance, then $\langle e^{\alpha x}\rangle_x = e^{\frac 12 \alpha^2}$, and therefore
$$
(1)\equiv
\prod_\mu\left\langle
e^{i\epsilon \sum_{a}\hat\Delta_\mu^a x_\mu}
\right\rangle_{x_\mu}=\exp\left[-\frac 12 \epsilon^2\sum_\mu\Big(\sum_a\hat\Delta_\mu^a\Big)^2\right].
$$
To compute the second term, recall that $\sigma_\mu^i$ is either 1 or 0 with probabilities $p$ and $1-p$:
$$
\begin{aligned}
(2)&=\prod_{i,\mu}\left\langle
e^{\frac iN \sum_{a}\hat\Delta_\mu^a g_i^a\sigma_\mu^i}
\right\rangle_{\sigma_\mu^i}\\
&=\prod_{i,\mu}\left(
(1-p)+p\, e^{\frac iN \sum_{a}\hat\Delta_\mu^a g_i^a}
\right)\\
&=\prod_{i,\mu}\left(
1+p \left[\frac {i}{N} \sum_{a}\hat\Delta_\mu^a g_i^a\right]
+\frac p2 \left[\frac i{N} \sum_{a}\hat\Delta_\mu^a g_i^a\right]^2+o(1/N^2)
\right)\\
&=\exp\left[
\frac{ip}{N} \sum_{i,\mu,a}\hat\Delta_\mu^a g_i^a
-\frac {p(1-p)}{2N^2}\sum_{i,\mu}\Big(\sum_a\hat\Delta_\mu^a g_i^a\Big)^2+o(1/N^2)
\right]
\end{aligned}
$$
Here we used $1+p\epsilon+\frac{p\epsilon^2}2+\dots=\exp(p\epsilon+p(1-p)\epsilon^2/2+\dots)$. Putting everything together:
\begin{multline*}
\left\langle Z^n\right\rangle_\mathrm{disorder} =
\int\!\prod_{i,a}\frac{d g_i^a}{N} e^{\beta\sum_{a}\tilde F(\{g_i^a\})}
\prod_{\mu,a}\frac{d\Delta_\mu^a\, d\hat\Delta_\mu^a}{2\pi}\, \theta(-\Delta_\mu^a)\\
\times \exp\left\{i\sum_{\mu,a}\hat\Delta_\mu^a \left(\Delta_\mu^a+\frac p{N}\sum_i g_i^a\right)-\frac12 \epsilon^2\sum_\mu\big(\sum_a\hat\Delta_\mu^a\big)^2-
\frac {p(1-p)}{2N^2}\sum_{i,\mu}\Big(\sum_a\hat\Delta_\mu^ag_i^a\Big)^2\right\}
\end{multline*}

\subsection{Decoupling indices $i$ and $\mu$}
In order to make progress, we need to eliminate terms that directly couple indices $\mu$ and $i$. To achieve this, we introduce a yet another set of variables, using the same trick of inserting a delta-function into our integral:
$$
\begin{aligned}
\text{Introduce }m^a&\equiv \frac 1{N}\sum_i g_i^a &&\Rightarrow &&\text{insert }
    1=\int\prod_a\frac{dm^a\,d\hat m^a}{2\pi}e^{i\hat m^a\left(m^a-\frac 1{N}\sum_i g_i^a\right)}\\
\text{Introduce }q^{ab}&\equiv \frac 1{N^2}\sum_i g_i^a g_i^b &&\Rightarrow &&\text{insert }
    1=\int\prod_{a\le b}\frac{dq^{ab}\,d\hat q^{ab}}{2\pi}e^{i\hat q^{ab}\left(q^{ab}-\frac 1{N^2}\sum_i g_i^a g_i^b\right)}\\
\end{aligned}
$$
These auxiliary variables will become our order parameters, as they capture the mean and variance of the deviation of resource availability from 1. The indices $i$ and $\mu$ are now decoupled, and we can split the integral accordingly. Recall that we have index $i$ labels $N$ different resources, while $\mu$ labels $P$ different strategies. Recall also that $\tilde F$ is a sum over $N$ terms:  $\tilde F\equiv \sum_{i=1}^N\tilde F_i$.
$$
\begin{aligned}
\left\langle Z^n\right\rangle
&=\int\prod_{a\le b}\frac{dq^{ab}\,d\hat q^{ab}}{2\pi}
\int\prod_{a}\frac{dm^{a}\,d\hat m^{a}}{2\pi}
\,\exp\left[i\sum_{a\le b}q^{ab}\hat q^{ab}+i\sum_a \hat m^a m^a\right]\\
&\times\prod_i\left\{
\int_{-\infty}^{N}\prod_a \frac{dg_i^a}{N}\ \exp\left[\sum_a\beta \tilde F_i(\{g_i^a\})-\frac i{N}\sum_a\hat m^a g_i^a-\frac i{N^2}\sum_{a\le b}\hat q^{ab}g_i^a g_i^b\right]\right\}\\
&\times\prod_\mu\left\{
\int\prod_a \frac{d\Delta_\mu^a\,d\hat\Delta_\mu^a}{2\pi}\prod_a\theta(-\Delta_\mu^a) \exp\left[i\sum_a\hat\Delta_\mu^a(\Delta_\mu^a+pm^a)-
\frac12\sum_{a,b}\left(p(1-p)q^{ab}+\epsilon^2\right)\hat \Delta_\mu^a \hat \Delta_\mu^b\right]\right\}\\
\\
&=
\int\prod_{a\le b}\frac{dq^{ab}\,d\hat q^{ab}}{2\pi}
\int\prod_{a}\frac{dm^{a}\,d\hat m^{a}}{2\pi}\exp\left[i\sum_{a\le b}q^{ab}\hat q^{ab}+i\sum_a \hat m^a m^a\right]
\times \prod_{i=1}^N A_i \times B^P,
\end{aligned}
$$
with $A_i$ and $B$ given by:
$$
\begin{aligned}
A_i&=\int_{-\infty}^{N}\prod_a \frac{dg^a}{N}\ \exp\left[\sum_a\beta \tilde F_i(g^a)-\frac {i}{N}\sum_a\hat m^a g^a-\frac i{N^2}\sum_{a\le b}\hat q^{ab}g^a g^b\right]\\
B&=\int\prod_a \frac{d\Delta^a\,d\hat\Delta^a}{2\pi}\prod_a\theta(-\Delta^a) \exp\left[i\sum_a\hat\Delta^a(\Delta^a+pm^a)-
\frac12\sum_{a,b}\left(p(1-p)q^{ab}+\epsilon^2\right)\hat \Delta^a \hat \Delta^b\right]
\end{aligned}
$$
After averaging over disorder, the problem becomes fully symmetric in indices $\mu$ (all strategies are drawn from the same distribution, so there is no inherent difference in how they contribute). The same would be true for resources, except in the interest of generality, we allowed the supply functions $H_i$ to be different for different resources.

\subsection{Decoupling replicas: the replica-symmetric ansatz}
The idea now is to treat the integrals over $m$ and $q$ as saddle-point integrals. We will shortly introduce a rescaling of variables that will make $\beta$ appear in the exponent to serve as the large parameter, making the saddle-point approximation appropriate. In this approximation, the integral is replaced by the value of the integrand at one location, the saddle point (up to a multiplicative prefactor, which, as explained below, is irrelevant for our purposes). When looking for the saddle point, we will make the assumption that it is symmetric under a permutation of replicas. This is the so-called ``replica-symmetric ansatz''. The validity of this assumption will be justified \textit{a posteriori} by the fact that the saddle-point we will find is ``well-behaved'' and the analytical results match the numerical simulations.

At a fully replica-symmetric saddle point, all components of $m^a$ must coincide. As for the matrix $q^{ab}$, all its diagonal elements must be equal, and all of the off-diagonal ones must be equal as well. The same holds for the conjugate variables $\hat m^a$, $\hat q^{ab}$. We therefore look for a saddle point of the following form:
$$
\hat q^{ab}=\left\{\begin{aligned}
&\hat q_D&&\text{ if $a=b$}\\
&\hat q_O&&\text{ if $a\neq b$}
\end{aligned}\right.,\qquad\qquad \hat m^a=\hat m^*,
$$
and similarly for $q^{ab}$ and $m^a$. With these assumptions:
$$
\log \langle Z^n\rangle =
\mathrm{extr}\;
\Big\{in\, q_D\hat q_D + i\frac{n(n-1)}{2}q_O\hat q_O+in \hat m^*m^*+ \sum_i\log A_i +P\log B\Big\}.
$$
(Why did we take the logarithm? Recall from Eq.~\ref{eq:trick} that our ultimate goal is to compute $Z^n-1$ in the $n\rightarrow0$ limit. Conveniently, in this limit, subtracting 1 is the same as taking the logarithm.) We stress that all expressions need only be computed to the leading exponential order. In particular, multiplicative constants are irrelevant as they only amount to an additive constant under the logarithm. Below, such constants will be omitted, and the ``equal'' signs will mean ``up to a constant multiplicative factor''.

\subsection{The limit $n\rightarrow 0$}
Note that in the expression we just found, $n$ enters in a way that allows taking the formal limit $n\rightarrow0$, which is our final goal. It turns out that this limit also makes the expressions for $\log A_i$ and $\log B$ somewhat easier to compute. Therefore, we write:
$$
\lim_{n\rightarrow0}\frac{\log \langle Z^n\rangle}n =
\lim_{n\rightarrow0}\mathrm{extr}\;
\Big\{i\, q_D\hat q_D - \frac i2 q_O\hat q_O+i \hat m^*m^*+ \frac 1n\sum_i\log A_i +\frac Pn\log B\Big\}.
$$
For reasons that will become clear later, we note that this can also be written as
\begin{equation}\label{eq:extr}
\lim_{n\rightarrow0}\frac{\log \langle Z^n\rangle}n =
\lim_{n\rightarrow0}\extr \Big\{i\left(\hat q_D-\frac 12 \hat q_O\right)q_D-\frac {q_D-q_O}2 (-i\hat q_O) + i\hat m^* m^*+ \frac 1n\sum_i\log A_i +\frac Pn\log B\Big\}
\end{equation}

\subsection{Computing $\log A_i$}
Recall the expression we denoted $A_i$:
$$
A_i=\int_{-\infty}^{N}\prod_a \frac{dg^a}{N}\ \exp\left[\sum_a\beta \tilde F_i(g^a)-\frac i{N}\sum_a\hat m^a g^a-\frac i{N^2}\sum_{a\le b}\hat q^{ab}g^a g^b\right]
$$
We first write:
$$
\begin{aligned}
\sum_{a\le b}\hat q^{ab}g^a g^b
&=\hat q_D\sum_a(g_a)^2+\frac 12 \hat q_O \sum_{a\neq b}g^ag^b\\
&=\hat q_D\sum_a(g_a)^2+\frac 12 \hat q_O \left[\Big(\sum_a g_a\Big)^2-\sum_a \big(g_a\big)^2\right]\\
&=\left(\hat q_D-\frac 12\hat q_O\right) \sum_a(g_a)^2+\frac 12 \hat q_O \Big(\sum_a g_a\Big)^2
\end{aligned}
$$
Now use Feynman's trick of removing the square by introducing an extra Gaussian variable:
$$\exp\left(\frac 12 Cx^2\right)=\int\mathcal Dz\, e^{z\,x\sqrt C}$$
(the curly $\mathcal D$ denotes the standard Gaussian measure with variance 1). This lets us write:
$$
\exp\left[-\frac i{N^2}\sum_{a\le b}\hat q^{ab}g^a g^b\right]=
\int\mathcal Dz\,\exp\left[z\frac{\sqrt{- i\hat q_O}}{N}\sum_a g^a-\frac i{N^2}\left(\hat q_D-\frac 12\hat q_O\right)\sum_a(g^a)^2\right].
$$
At the price of introducing an extra Gaussian variable, all replicas are now fully decoupled. Plugging this into our expression for $A_i$:
$$
A_i=\int \mathcal{D}z \left[
\int_{-\infty}^N \!\frac{dg}N\, \exp \left(\beta \tilde F_i(g)
- \frac{i}{N^2}\left(\hat q_D-\frac 12 \hat q_O\right) g^2 -\frac 1N \left(i \hat m^* - z\sqrt{-i\hat q_O}\right) g
\right)
\right]^n
$$
Conveniently, for small $n$:
$$
\log\int\mathcal\!Dz\, x^n=\log\left[\int\!\mathcal Dz\, (1+n\log x+\dots)\right]=\log \left[1+n\int\!\mathcal Dz\,\log x+\dots\right]=
n\int\!\mathcal Dz\,\log x+\dots
$$
Therefore:
$$
\lim_{n\rightarrow0}\frac {\log A_i}n =
\int\! \mathcal{D}z\, \log
\int_{-\infty}^N \!\frac{dg}N\, \exp \left[\beta \tilde F_i(g)
- \frac{i}{N^2}\left(\hat q_D-\frac 12 \hat q_O\right) g^2 -\frac 1N \left(i \hat m^* - z\sqrt{-i\hat q_O}\right) g
\right]
$$
Introduce rescaled variables as follows:
$$
\begin{aligned}
i\left(\hat q_D-\frac 12 \hat q_O\right)&\equiv \beta a\\
\sqrt{-i\hat q_O}&\equiv \frac{\beta b}{\sqrt N}\\
i\hat m^* &\equiv \beta \hat m\\
\end{aligned}
$$
In the new variables:
$$
\lim_{n\rightarrow0}\frac {\log A_i}n =
\int\! \mathcal{D}z\, \log
\int_{-\infty}^N \!\frac{dg}N\, \exp \beta\left[ \tilde F_i(g)
- a\frac{g^2}{N^2} - \left(\hat m-\frac {zb}{\sqrt{N}}\right)\frac gN \right]
$$
Substitute the general form of $\tilde F_i$ from~\eqref{eq:Fgeneral}, for a close-to-symmetric resource supply~\eqref{eq:symmetricAnsatz}:
$$
\lim_{n\rightarrow0}\frac {\log A_i}n =
\int\! \mathcal{D}z\, \log
\int_{-\infty}^N \!\frac{dg}N\, \exp \beta\left[-\left(a+\frac 1{2\gamma_i}\right)\left(\frac {g}N\right)^2
+\left(-\bar\tau-\hat m+\frac {zb+\delta\tau_i}{\sqrt N}\right)\frac gN \right]
$$
In the limit $\beta\rightarrow\infty$:
$$
\lim_{n\rightarrow0}\frac {\log A_i}n =
\beta\int\! \mathcal{D}z\, \max_{y<1} \left[-\left(a+\frac 1{2\gamma_i}\right)y^2
+\left(-\bar\tau-\hat m+\frac {zb+\delta\tau_i}{\sqrt N}\right)y \right]
$$
Here $y\equiv \frac gN$. The quadratic form $-P y^2+Qy$ is maximized at $y=\frac{Q}{2P}$, reaching the maximal value of $\frac{Q^2}{4P}$. In our case, therefore, for a given $z$ the quadratic form reaches its maximum at
$$
y_i^*(z)=\frac{-\bar\tau-\hat m+(zb+\delta\tau_i)/{\sqrt N}}{2a+1/\gamma_i}.
$$
Let us shift the variable $\hat m$ by defining $\hat m\equiv-\bar\tau-\frac{\delta\hat m}{\sqrt N}$. For now, we can treat this as a simple change of variables. We find:
$$
y_i^*(z)=\frac 1{\sqrt N} \frac{zb+\delta\tau_i+\delta\hat m}{2a+1/\gamma_i} = \frac 1{\sqrt N} \frac{zb+\delta\tau_i+\delta\hat m}{2a+1/\bar\gamma}
+o\left(\frac{1}{\sqrt N}\right).
$$
Recall that $\hat m$ is one of the variables over which the extremum is computed in~\eqref{eq:extr}. The shift of $\hat m$ that we just did constitutes an assumption, namely that the extremum is located close to $\hat m\approx\bar\tau$ (i.e.\ that the difference is at most of order $1/\sqrt{N}$). We will check the consistency of this assumption below. Note that if we are correct to assume this, then $y_i^*(z)$ is small, justifying the expansion of resource depletion functions to first order in $y$ .

The integral over $z$ is now simple to compute:
$$
\lim_{n\rightarrow0}\frac {\log A_i}n =
\frac \beta{2N}\int\! \mathcal{D}z\, \frac{(zb+\delta\tau_i+\delta\hat m)^2}{2a+1/\bar\gamma}=
\frac \beta{2N}\frac{b^2+(\delta\tau_i+\delta\hat m)^2}{2a+1/\bar\gamma}
$$
Finally, performing the sum over $i$ and recalling that $\sum_i\delta\tau_i=0$, we find a very simple final expression:
$$
\sum_i\lim_{n\rightarrow0}\frac {\log A_i}n =
\beta\frac{b^2+\delta\hat m^2+\overline{\delta \tau^2}}{4a+2/\bar\gamma},
$$
where $\overline{\delta \tau^2}\equiv \frac 1N\sum_i(\delta\tau_i)^2$ is the variance of resource supply across $i$.

\subsection{Computing $\log B$}
Recall the definition of $B$:
$$
B=\int\prod_a \frac{d\Delta^a\,d\hat\Delta^a}{2\pi}\prod_a\theta(-\Delta^a) \exp\left[i\sum_a\hat\Delta^a(\Delta^a+pm^a)-
\frac12\sum_{a,b}\left(p(1-p)q^{ab}+\epsilon^2\right)\hat \Delta^a \hat \Delta^b\right]
$$
Proceeding as above, we decompose
$$
-\sum_{a,b}\frac{p(1-p)q^{ab}+\epsilon^2}2 \hat\Delta^a\hat\Delta^b=-\frac{p(1-p)}2(q_D-q_O)
\sum_a (\hat\Delta^a)^2-\frac {p(1-p)q_O+\epsilon^2}2\Big(\sum_a\hat\Delta^a\Big)^2.
$$
We then remove the square by introducing an extra Gaussian variable, making all replicas fully decoupled:
$$
B=\int\!\mathcal Dw\,\left[
\int\frac{d\Delta\,d\hat\Delta}{2\pi}\theta(-\Delta) \exp\left(i\hat\Delta(\Delta+pm^*)-
\frac{p(1-p)}2(q_D-q_O)\hat\Delta^2+iw\hat\Delta\sqrt{p(1-p)q_O+\epsilon^2}\right)
\right]^n.
$$
Note that the integral over $\hat \Delta$ inside the square brackets is a simple Gaussian integral, and we can write:
\begin{equation}\label{eq:Bintermediate}
B=\int\mathcal Dw \left[
\int_{-\infty}^0\! \frac{d\Delta}{\sqrt{2\pi p(1-p)(q_D-q_O)}}\,
\exp\left\{-\frac12 \frac{\left(\Delta+pm^*+w\sqrt{p(1-p)q_O+\epsilon^2}\right)^2}{p(1-p)(q_D-q_O)}\right\}
\right]^n\\
\end{equation}
Introduce a notation $E(x)\equiv\int_x^\infty \frac {dy}{\sqrt{2\pi}}e^{-y^2/2}$. This is essentially the error function, up to a couple constants that would be a nuisance to carry around: $E(x)=\frac 12\erfc\frac{x}{\sqrt2}$. We can then write:
$$
\lim_{n\rightarrow0}\frac {\log B}n=\int\!\mathcal Dw\, \log E\left[
-\frac{pm^*+w\sqrt{p(1-p)q_O+\epsilon^2}}{\sqrt{p(1-p)(q_D-q_O)}}
\right].
$$
As above for $A$, we now introduce a rescaled variable $x$, and some convenient notations:
$$
\begin{aligned}
q_D-q_O &\equiv \frac{Nx}{\beta}\\
q_D\approx q_O &\equiv q\\
\sqrt{p(1-p)q+\epsilon^2}&\equiv \psi\\
pm^*/\psi&\equiv \lambda
\end{aligned}
$$
In the new variables:
$$
\lim_{n\rightarrow0}\frac {\log B}n=\int\!\mathcal Dw\, \log E\left[
-\sqrt{\frac\beta N}\frac{(w+\lambda)\psi}{\sqrt{p(1-p)x}}
\right].
$$
The logarithm of $E(x)\equiv\int_x^\infty \frac {dy}{\sqrt{2\pi}}e^{-y^2/2}$ in the large-argument limit is very simple. Indeed:
$$
\lim_{\beta\rightarrow\infty} E(\sqrt\beta x)\simeq\left\{
\begin{aligned}
&1-C\exp(-\beta x^2/2) &&\text{if $x<0$}\\
&C\exp(-\beta x^2/2) &&\text{if $x>0$}\\
\end{aligned}
\right.
$$
Therefore (omitting additive constants as always):
$$
\lim_{\beta\rightarrow\infty} \log E(\sqrt\beta x)\simeq\left\{
\begin{aligned}
&0 &&\text{if $x<0$}\\
&-\beta x^2/2 &&\text{if $x>0$}\\
\end{aligned}
\right.
$$
Plugging this into our expression for $\log B$, we find
$$
\lim_{n\rightarrow0}\frac {\log B}n=
-\frac {\beta\psi^2}{2Np(1-p)x} \int_{-\infty}^{-\lambda}\!\mathcal Dw\, (w+\lambda)^2
=-\frac {\beta\psi^2}{2Np(1-p)x} I(\lambda),
$$
where $I(\lambda)$ can be expressed in terms of the error function:
$$
I(\lambda)\equiv\int_0^\infty e^{-\frac{(w-\lambda)^2}2}w^2\frac{dw}{\sqrt{2\pi}} = -\frac{\lambda}{\sqrt{2\pi}}e^{-\frac{\lambda^2}2}+\frac {1+\lambda^2}2\erfc\left(\frac \lambda{\sqrt2}\right).
$$

\subsection{Putting everything together}
Combining the results above, plugging them into~\eqref{eq:extr}, and recalling that in the large-$N$ limit, $P$ also goes to infinity with $\frac PN\equiv \alpha$, we find:
$$
\begin{aligned}
\langle \log Z\rangle
&=\lim_{n\rightarrow0}\frac{\langle Z^n-1\rangle}n =\lim_{n\rightarrow0}\frac{\log \langle Z^n\rangle}n \\
&=\lim_{n\rightarrow0}\extr \Big\{i\left(\hat q_D-\frac 12 \hat q_O\right)q_D-\frac {q_D-q_O}2 (-i\hat q_O) + i\hat m^* m^*+ \frac 1n\sum_i\log A_i +\frac Pn\log B\Big\}\\
&=\beta \extr \Big\{aq-\frac{b^2x}2 + \left(-\bar\tau-\frac{\delta\hat m}{N}\right)\frac{\psi\lambda}{p}
+\frac{b^2+\delta\hat m^2+\overline{\delta \tau^2}}{4a+2/\bar\gamma}
-\frac{\alpha\psi^2}{2p(1-p)x}I(\lambda)\Big\}.
\end{aligned}
$$
Recall that $\psi\equiv\sqrt{p(1-p)q+\epsilon^2}$, so that the extremum is taken over six variables: $\delta\hat m$, $a$, $b$, $q$, $\lambda$ and $x$. Consider the extremum condition for $\delta\hat m$:
$$
\frac{\delta\hat m}{2a+1/\bar\gamma}=\frac 1N\frac{\psi\lambda}p.
$$
As $N\rightarrow\infty$, we therefore have $\delta\hat m\rightarrow0$, demonstrating that the approximation $\hat m\approx\bar\tau$ was indeed self-consistent. Setting $\delta\hat m=0$, we find:
$$
\begin{aligned}
\langle \log Z\rangle
=\beta \extr \Big\{aq-\frac{b^2x}2 - \bar\tau\frac{\psi\lambda}{p}
+\frac{b^2+\overline{\delta \tau^2}}{4a+2/\bar\gamma}
-\frac{\alpha\psi^2}{2p(1-p)x}I(\lambda)\Big\}.
\end{aligned}
$$
The powers of $N$ and $\beta$ in our rescaled variables were chosen to ensure that this expression no longer depends on $N$ and is proportional to $\beta$. Conveniently, the extremum conditions for variables $a$ and $b$ can also be solved, and these variables eliminated:
$$
\begin{aligned}
a &= \frac{\bar\gamma-x}{2x\gamma}\\
b^2 &= \frac{q}{x^2}-\overline{\delta \tau^2}
\end{aligned}
$$
Our final expression for the partition function:
\begin{equation}\label{eq:smry:Z1}
\langle \log Z\rangle
=\beta \extr \Big\{\frac{\bar\gamma-x}{2x\bar\gamma}q + \frac{\overline{\delta \tau^2}}2x - \frac{\lambda\tau\psi(q)}{p}-\frac{\alpha\psi^2(q)}{2xp(1-p)}I(\lambda)\Big\}
\end{equation}
The extremum is to be computed over $q$, $x$ and $\lambda$. Here $\alpha$, $p$ and $\epsilon$ (hidden in $\psi\equiv\sqrt{p(1-p)q+\epsilon^2}$) are parameters characterizing the pool of competitors (number of strategies, typical functional sparsity, and intrinsic scatter cost, respectively). Parameters $\bar\tau$, $\overline{\delta \tau^2}$ and $\bar \gamma$ characterize resource supply (respectively: average capacity, variability across resources, and average ``elasticity''). For the simplest single-parameter resource model used in the text, the influx of resource $i$ is fixed at $R_i$, whose average, without loss of generality, can be set to 1. In this case we have $\bar\tau=1$, $\bar \gamma=1$ and the remaining parameter is the variance of resource supply $\overline{\delta \tau^2}$, denoted $\dRSq$ in the main text.

\section{The saddle-point equations}
\subsection{Simplifying the equations to solve them numerically}
Hidden in $I(\lambda)$ is the error function $\erfc$, which means that the extremum of~\eqref{eq:smry:Z1} cannot be found analytically. However, the equations can be simplified to a form where they can either be solved numerically, or investigated analytically in certain limits.

The extremum conditions:
\begin{align}
\frac{\partial}{\partial q}:\qquad& \frac{1-\alpha I(\lambda)}x = \frac{1}{\bar\gamma}+\frac{(1-p)\lambda\bar\tau}{\psi}\label{eq:first}\\
\frac{\partial}{\partial x}:\qquad& \dtSq - \frac{q}{x^2}+\frac{\alpha\psi^2}{x^2p(1-p)} I(\lambda)=0\label{eq:second}\\
\frac{\partial}{\partial \lambda}:\qquad& -2-\frac{\alpha\psi}{x\bar\tau(1-p)}\frac{dI}{d\lambda}=0\nonumber
\end{align}
Using the first and the third equations, we write:
$$
\frac{1-\alpha I(\lambda)}{\frac 1{\bar\gamma}+\frac{(1-p)\lambda\bar\tau}{\psi}} = x = -\frac{\alpha\psi}{2\bar\tau(1-p)}\frac{dI}{d\lambda}
$$
Rearranging, we find a way to express $\psi$ (and thus $q$) in terms of $\lambda$ only:
\begin{equation}\label{eq:psi}
\frac{\psi}{\bar\gamma\bar\tau} = \frac{2(1-p)(1-\alpha I(\lambda))}{-\alpha I'(\lambda)}-(1-p)\lambda
\end{equation}
Plugging this into the first equation, and recalling the definition of $I(\lambda)$, we find a very simple expression for $x$:
\begin{equation}\label{eq:x}
\frac{x}{\bar\gamma}=1-\alpha I(\lambda)+\frac{\alpha\lambda}{2} \frac{dI}{d\lambda} = 1-\alpha E(\lambda).
\end{equation}
(Recall that $E(\lambda)\equiv \frac12\erfc(\lambda/\sqrt 2)$.) This result makes it possible to eliminate $x$ from the equations. Plugging all this into~\eqref{eq:second}, we find an equation that involves $\lambda$ only, and can easily be solved numerically. Once $\lambda$ is known, equations~\eqref{eq:psi} and~\eqref{eq:x} determine $q$ and $x$.

\subsection{Investigating the limit $\epsilon\rightarrow0$}
To study the equations analytically in the limit $\epsilon\rightarrow0$, we plug the expression for $x$ into~\eqref{eq:first}, and reorganize the terms in equation~\eqref{eq:second}, putting it into the form cited in the main text:
\begin{equation*}
\left\{\begin{aligned}
&\frac{1-\alpha I(\lambda)}{1-\alpha E(\lambda)} = 1+\frac{\lambda}{\psi}\,(1-p)\bar\gamma \\ 
&\psi^2(1-\alpha I(\lambda))=\epsilon^2+\big(1-\alpha E(\lambda)\big)^2\, \bar\gamma^2p(1-p)\dtSq
\end{aligned}\right.
\end{equation*}

The easiest way to derive the expression for the critical line given in the main text is to observe the following. At $\epsilon=0$, we see that simultaneously setting $\psi=0$ and $1-\alpha E(\lambda)=0$ yields a solution. One can check that in the vicinity of the transition, both go to zero linearly, so that their ratio remains well-defined. Omitting the negligible first terms in the right-hand sides of both equations, we rewrite them as follows:
\begin{equation*}
\left\{\begin{aligned}
&\frac\psi{\bar\gamma}\frac{1-\alpha I(\lambda)}{1-\alpha E(\lambda)} = \lambda\,(1-p)\\ 
&\frac{\psi^2}{\bar\gamma^2}\frac{(1-\alpha I(\lambda))^2}{(1-\alpha E(\lambda))^2}=p(1-p)\dtSq (1-\alpha I(\lambda))
\end{aligned}\right.
\end{equation*}

At the critical line, the system is degenerate, i.e.\ the two equations are proportional. We immediately read off the condition that must be satisfied:
$$
p(1-p)\dtSq_{\text{crit}} (1-\alpha_{\text{crit}} I(\lambda)) = \lambda^2\,(1-p)^2 \qquad \Rightarrow \qquad
\dtSq_{\text{crit}}  = \frac{1-p}p\frac{\lambda^2}{1-\alpha_{\text{crit}} I(\lambda)}.
$$
This is the expression quoted in the main text (with $\lambda$ being fixed by the condition $1-\alpha_{\text{crit}}E(\lambda)=0$).

\section{Computing the number of survivors at equilibrium}
To find the number of species that survive at equilibrium in our model, we set out to compute the distribution of the observable $\Delta$. Recall that $\Delta$ is the ``resource surplus''; for a given set of competitors, there is a discrete set of values of $\Delta_\mu$, the resource surplus experienced by each species. However, after we average our partition function over disorder, $\Delta$ becomes a random variable, drawn out of a certain distribution, whose shape is encoded in the partition function. This distribution is what we now set out to compute.

The $\Delta$-dependent part of the partition function is fully contained in the expression for $B$. Recall the intermediate expression~\eqref{eq:Bintermediate} derived earlier, when computing $B$:
$$
\begin{aligned}
B&=\int\mathcal Dw \left[
\int_{-\infty}^0\! \frac{d\Delta}{\sqrt{2\pi p(1-p)\,Nx/\beta}}\,
\exp\left\{-\frac\beta2 \frac{\left(\Delta+pm+w\psi\right)^2}{p(1-p)\,Nx}\right\}
\right]^n\\
&\equiv\int\mathcal Dw \left[
\int_{-\infty}^0\! \frac{d\Delta}{\sqrt{2\pi/\beta'}}\,
\exp\left\{-\frac{\beta'}{2}\left(\Delta+pm+w\psi\right)^2\right\}
\right]^n,
\end{aligned}
$$
Here we introduced $\beta'\equiv\frac{\beta}{p(1-p)Nx}$ to make the notations slightly less heavy. Let us put this expression in the following form, retaining one copy of the integral over $\Delta$, while evaluating the remaining $n-1$ copies as before:
$$
B = \int_{-\infty}^0 \frac{d\Delta}{\sqrt{2\pi/\beta'}}\int\!\mathcal Dw\,
\exp\left(-\frac {\beta'}2 (\Delta+pm+w\psi)^2\right)
\left[E\left(-\sqrt{\beta'}(w\psi+pm)\right)\right]^{n-1}
$$
Once again, $E(x)$ is a short-hand for $\frac12\erfc(x/\sqrt 2)$. Sending $n\rightarrow 0$ and recalling the notation $\lambda\equiv \frac{pm}{\psi}$:
$$
B = \int_{-\infty}^0 \frac{d\Delta}{\sqrt{2\pi/\beta'}}\int\!\mathcal Dw\,
\exp\left\{-\frac {\beta'}2 (\Delta+(w+\lambda)\psi)^2\right\}
\frac 1{E\left(-\sqrt{\beta'}\,\psi(w+\lambda)\right)}.
$$
From this we infer the distribution of $\Delta$ for $\Delta\le0$ (positive $\Delta$ are forbidden):
$$
p(\Delta) = \int\!\mathcal Dw\,\frac{1}{\sqrt{2\pi/\beta'}}
\frac {\exp\left\{-\frac {\beta'}2 (\Delta+(w+\lambda)\psi)^2\right\}}{E\left(-\sqrt{\beta'}\,\psi(w+\lambda)\right)}
$$
This is a complicated-looking expression, but its $\beta'\rightarrow\infty$ limit can in fact be computed very easily, using the following trick. By definition of the function $E(x)$, this expression can be rewritten as follows:
$$
p(\Delta) = \int\!\mathcal Dw\,\left[
\frac {\frac{1}{\sqrt{2\pi/\beta'}}\exp\left\{-\frac {\beta'}2 (\Delta+(w+\lambda)\psi)^2\right\}}
{\int_{-\infty}^0\!\frac{dy}{\sqrt{2\pi/\beta'}}\,\exp\left\{-\frac {\beta'}2 (y+(w+\lambda)\psi)^2\right\}}\right]
\equiv \int\!\mathcal Dw\,p(\Delta|w).
$$

The key observation is that here, the ``conditional distribution'' $p(\Delta|w)$ is (by inspection) a properly normalized distribution, for all $w$:
\begin{equation}\label{eq:normCond}
\forall w:\:\int_0^\infty\!d\Delta\,p(\Delta|w)=1
\end{equation}
We now compute the limit $\lim_{\beta'\rightarrow\infty}p(\Delta|w)$. For $w>-\lambda$, this is a Gaussian distribution of width $\frac 1{\sqrt{\beta'}}\rightarrow 0$ centered at $\Delta=-\psi(\lambda+w)<0$. For $w<-\lambda$, the probability density is highest at $\Delta=0$ and goes to zero everywhere else (recall that positive $\Delta$ are forbidden). Therefore, the normalization condition~\eqref{eq:normCond} immediately tells us that:
$$
\lim_{\beta'\rightarrow\infty}p(\Delta|w)=\left\{
\begin{aligned}
&\delta(\Delta+\psi(\lambda+w))&&\text{if $w>-\lambda$}\\
&\delta(\Delta)&&\text{if $w<-\lambda$}\\
\end{aligned}
\right.
$$
As a result:
\begin{align}
p(\Delta)&=\int_{-\lambda}^{\infty}\frac{dw}{\sqrt{2\pi}}e^{-\frac{w^2}2}\delta(\Delta+\psi(\lambda+w))+
\int_{-\infty}^{-\lambda}\frac{dw}{\sqrt{2\pi}}e^{-\frac{w^2}2}\delta(\Delta)\nonumber\\
&=\frac{1}{\sqrt{2\pi\psi^2}}e^{-\frac{(\Delta+\lambda\psi)^2}{2\psi^2}}\cdot\theta(-\Delta)+E(\lambda)\delta(\Delta)\label{eq:SingleReplicaDeltaDistrib}.
\end{align}
This is the expression quoted in the main text. The weight of the delta-shaped peak at $\Delta=0$ corresponds to the species whose resource balance is met; these are the species who survive competition.

\section{Validation of analytical results against simulations}
Numerical simulations were performed in MatLab; a script generating all figures is available as Supplementary File~1. Random instances were generated by randomly drawing strategy vectors and costs as described in the main text. All simulations used $N=50$ for the number of resources. The equilibrium of a community was determined by a direct $N$-dimensional numerical optimization of $\tilde F=\sum_i R_i/T_i$ subject to the linear constraints defining the ``unsustainable region'' $\Omega$. The uneven resource supply was always implemented as a bimodal distribution shown in Fig.~\ref{fig:cross}B (upper panel); the amplitude of the step was adjusted to match the required magnitude of $\dRSq$.

\subsection{Figure~\ref{fig:phases}B}
500 simulations were performed at $\alpha=10$, $\epsilon=10^{-3}$ and $\dRSq=1$ (reusing the dataset computed for Fig.~\ref{fig:cross}B). These parameters are comfortably in the $S$-phase, so the expected number of survivors is $N$. After numerical equilibration, all but the top $N$ values of the resource surplus $\Delta$ were recorded (the top $N$ are within numerical error of $0$ and correspond to survivors). Panel~\ref{fig:phases}B shows a joint histogram of these values recorded over 500 instances, appropriately normalized to be comparable with the theoretical distribution. For visualization purposes, the delta-shaped peak is shown as a rectangle, its height is meaningless. The theoretical prediction for the number of survivors is verified in panel~\ref{fig:phases}C.

\subsection{Figure~\ref{fig:phases}C}
The theoretical curves are overlaid with simulations obtained as follows. For 10 values of $\alpha$ equispaced between 2 and 5, and for three values of $\epsilon$ indicated on the plot (0.1, 0.03 and $10^{-4}$), communities were equilibrated using MatLab solver \texttt{fmincon} with numerical precision parameter $10^{-10}$. Species whose resource surplus was within $10^{-8}$ of zeros were declared as survivors (for comparison, the resource surplus of the first extinct species was typically $\approx 10^{-6}$). The panel shows the mean number of survivors over 500 instances for each $(\alpha,\epsilon)$ pair.

\subsection{Figure~\ref{fig:cross}A}
For each indicated epsilon, 10 simulations were performed for each $\alpha$ in the list: $\{1, 2, 3, 4, 5, 6, 8, 10, 15, 20, 50\}$.

\subsection{Figure~\ref{fig:cross}B}
For $\epsilon=10^{-3}$ and $\dRSq=1$, we performed 500 simulations at $\alpha=2$ for $V$ phase, and at $\alpha=10$ for S-phase. The latter dataset was also used for generating the resource surplus histogram shown in \ref{fig:phases}B.

\newpage
\section{Interpretation of the V-phase and the S-phase}
\begin{figure}[h!]
\centering
\includegraphics[width=\linewidth]{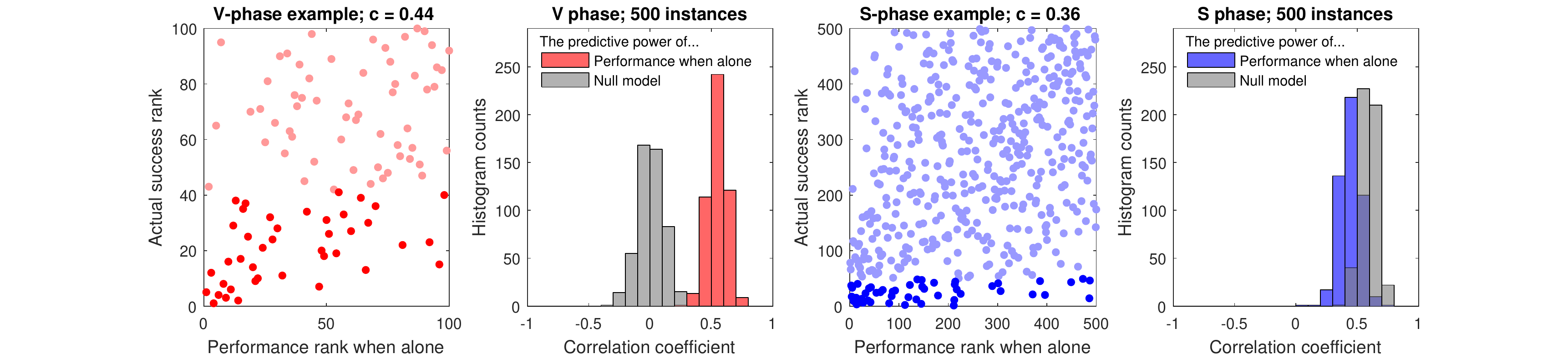}
\caption{Define a species ``success rank'' by ordering all survivors by decreasing abundance at equilibrium, followed by all the species that went extinct, in order of increasing resource insufficiency $|\Delta_\mu|$. Now, consider two quantities that could potentially be used to predict this success. To obtain the first, we measure the growth rate of each species when it is placed into the externally supplied conditions, with no competitors present. Alternatively, the ``null model'' performance predictor is simply the species' cost per pathway: clearly, the high-cost species are less likely to survive. We call the latter quantity the ``null model'', because it includes no information about the environment to which the community is subjected.\\
We now ask if either of the two quantities are in fact predictive of the true success of a species. We measure predictive power by the Spearman (rank-order) correlation coefficient between the predictor being tested and the true success rank. In the V-phase, the species' performance measured in external conditions is indeed predictive of the success rank (panel \textbf{A}.) A histogram of the correlation coefficients over 500 instances (panel \textbf{B}) confirms that this predictor significantly outperforms the null model. In contrast, in the S-phase, the environment-specific performance measured in the externally imposed conditions becomes irrelevant (panels \textbf{C} and \textbf{D}). Although the correlation observed in panel C is non-zero, this is due to the generic fact that low-cost organisms are generally more likely to survive that high-cost ones. This is demonstrated in panel D: unlike the more intuitive V-phase, in the S-phase the ``environment-aware'' predictor performs \textit{worse} than the null model.\label{fig:SXXphases}}
\end{figure}


\begin{thebibliography}{99}
%
\bibitem{Gill06}
S. R. Gill, M. Pop, R. T. DeBoy, P. B. Eckburg, P. J. Turnbaugh, B. S. Samuel, J. I. Gordon, D. A. Relman, C. M. Fraser-Liggett, and K. E. Nelson.
Metagenomic analysis of the human distal gut microbiome.
Science \textbf{312}, 5778 (2006).
%
\bibitem{Caporaso11}
J. G. Caporaso, C. L. Lauber, W. A. Walters, D. Berg-Lyons, C. A. Lozupone, P. J. Turnbaugh, N. Fierer, and R. Knight.
Global patterns of 16S rRNA diversity at a depth of millions of sequences per sample.
PNAS \textbf{108} (2011).
%
\bibitem{Lozupone12}
C. A. Lozupone, J. I. Stombaugh, J. I. Gordon, J. K. Jansson, and R. Knight.
Diversity, stability and resilience of the human gut microbiota.
Nature \textbf{489}, 7415 (2012).
%
\bibitem{HMP}
Human Microbiome Project Consortium.
Structure, function and diversity of the healthy human microbiome.
Nature \textbf{486}, 7402 (2012).
%
\bibitem{EMP}
J. A. Gilbert, J. K. Jansson, and R. Knight.
The Earth Microbiome Project: successes and aspirations.
BMC Biol \textbf{12} (2014).
%
\bibitem{Beiko15}
R. G. Beiko.
Microbial malaise: how can we classify the microbiome?
Trends Microbiol \textbf{23}, 11 (2015).
%
\bibitem{Fussmann07}
G. F. Fussmann, M. Loreau, and P. A. Abrams.
Eco-evolutionary dynamics of communities and ecosystems.
Functional Ecology \textbf{21}, 3 (2007).
%
\bibitem{Pelletier09}
F. Pelletier, D. Garant, and A. P. Hendry.
Eco-evolutionary dynamics.
Philosophical Transactions of the Royal Society B-Biological Sciences, \textbf{364}, 1523 (2009).
%
\bibitem{Henson15}
S. M. Henson, J. M. Cushing, and J. L. Hayward.
Introduction to special issue on eco-evolutionary dynamics.
Natural Resource Modeling \textbf{28}, 4 (2015).
%
\bibitem{Schoener11}
T. W. Schoener.
The newest synthesis: Understanding the interplay of evolutionary and ecological dynamics.
Science \textbf{331}, 6016 (2011).
%
\bibitem{Grover97}
J. P. Grover.
Resource competition.
Population and community biology series. Chapman \& Hall, London; New York, 1st edition, 1997.
%
\bibitem{ScottPhillips14}
T. C. Scott-Phillips, K. N. Laland, D. M. Shuker, T. E. Dickins, and S. A. West.
The niche construction perspective: A critical appraisal.
Evolution, \textbf{68}, 5 (2014).
%
\bibitem{McCook94}
L. J. McCook.
Understanding ecological community succession -- causal-models and theories, a review.
Vegetatio \textbf{110}, 2 (1994).
%
\bibitem{MacArthur}
R. MacArthur.
Species packing, and what interspecies competition minimizes.
PNAS \textbf{64}, 4 (1969).
%
\bibitem{Tilman82}
D. Tilman.
Resource competition and community structure.
Monogr Popul Biol \textbf{17} (1982).
%
\bibitem{CWC}
M. Tikhonov.
Community-level cohesion without cooperation.
eLife \textbf{5} (2016).
%
\bibitem{Fischbach07}
M. A. Fischbach and J. Clardy.
One pathway, many products.
Nature Chemical Biology \textbf{3}, 7 (2007).
%
\bibitem{Fischbach11}
M. A. Fischbach and J. L. Sonnenburg.
Eating for two: How metabolism establishes lnterspecies interactions in the gut.
Cell Host \& Microbe \textbf{10}, 4 (2011).
%
\bibitem{StatMechOfLearning}
A. Engel and C. van den Broeck.
Statistical mechanics of learning.
Cambridge, UK; New York, NY; Cambridge University Press. (2001).
%
\bibitem{Gardner88}
E. Gardner,
The space of interactions in neural network models.
J. Phys. A \textbf{21}, 257 (1988).
%
\bibitem{DeMartino10}
A. De Martino and E. Marinari.
The solution space of metabolic networks: Producibility, robustness and fluctuations.
Journal of Physics: Conference Series \textbf{233}, 1 (2010).
\end{thebibliography}
\end{document}